\documentclass[aps,prd,preprint,tightenlines,nofootinbib,showpacs]{revtex4}

\usepackage{epsfig}
\usepackage{graphicx}
\usepackage{dcolumn}
\usepackage{bm}

\def \bdir{{\cal B}({\rm dir})}
\def \bea{\begin{eqnarray}}
\def \beq{\begin{equation}}
\def \bhc{\brp\brh}
\def \br{{\cal B}}
\def \brp{{\cal B}_{\psi}}
\def \brh{{\cal B}_h}
\def \ec{\eta_c}
\def \eea{\end{eqnarray}}
\def \eeq{\end{equation}}
\def \ege1{E(\gamma_{\rm E1})}
\def \gec{\Gamma(\ec)}
\def \ggJ{\gamma \gamma}
\def \ghc{\Gamma(\hc)}
\def \gme1{\gamma_{\rm E1}}
\def \hc{h_c}
\def \jp{J/\psi}
\def \kl{K^0_L}
\def \ks{K^0_S}
\def \ls{\stackrel{<}{\sim}}
\def \mav{\langle M(1^3P) \rangle}
\def \mpla#1#2#3{Mod.\ Phys.\ Lett.\ A {\bf#1}, #2 (#3)}
\def \npbps#1#2#3{Nucl.\ Phys.\ B Proc.\ Suppl.\ {\bf#1}, #2 (#3)}
\def \plb#1#2#3{Phys.\ Lett.\ B {\bf#1}, #2 (#3)}
\def \pp{\psi(2S)}
\def \prd#1#2#3{Phys.\ Rev.\ D {\bf#1}, #2 (#3)}

\def \pz{\pi^0}

\newcommand{\kkpp}{K^{+}K^{-}\pi^{+}\pi^{-}}
\newcommand{\fourpi}{\pi^{+}\pi^{-}\pi^{+}\pi^{-}}
\newcommand{\kkpizero}{K^{+}K^{-}\pi^{0}}
\newcommand{\ppetaone}{\pi^{+}\pi^{-}\eta(\gamma\gamma)}
\newcommand{\ppetatwo}{\pi^{+}\pi^{-}\eta(\pi^{+}\pi^{-}\pi^{0})}
\newcommand{\direct}{\pp \to \gamma \ec}
\newcommand{\cascade}{\pp \to \pz \hc \to \pz (\gamma \ec)}

\newcommand{\MeV}{\rm MeV}

\newcommand{\kskp}{\ks K^{\pm}\pi^{\mp}}
\newcommand{\klkp}{\kl K^{\pm}\pi^{\mp}}
\newcommand{\sys}{\rm sys}
\newcommand{\stat}{\rm stat}

\begin{document}

\preprint{CLNS 05/1920}       
\preprint{CLEO 05-12}         

\title{Observation of the $^1P_1$ State of Charmonium}

\author{P.~Rubin}
\affiliation{George Mason University, Fairfax, Virginia 22030}
\author{C.~Cawlfield}
\author{B.~I.~Eisenstein}
\author{G.~D.~Gollin}
\author{I.~Karliner}
\author{D.~Kim}
\author{N.~Lowrey}
\author{P.~Naik}
\author{C.~Sedlack}
\author{M.~Selen}
\author{E.~J.~White}
\author{J.~Williams}
\author{J.~Wiss}
\affiliation{University of Illinois, Urbana-Champaign, Illinois 61801}
% ...
\author{(CLEO Collaboration)} 
\noaffiliation


\date{\today}

\begin{abstract} 

The spin-singlet P-wave state of charmonium, $\hc(^1P_1)$, has been observed in
the decay $\pp \to \pz \hc$ followed by $\hc \to \gamma \ec$.  Inclusive
and exclusive analyses of the $M(\hc)$ spectrum have been performed.  Two
complementary inclusive analyses select either a range of energies for the
photon emitted in $\hc \to \gamma \ec$ or a range of values of $M(\ec)$.  These
analyses, consistent with one another within statistics, yield $M(h_c) =[3524.9
\pm 0.7~{\rm (stat)} \pm 0.4~{\rm (sys)}]$ MeV/$c^2$ and a product of the
branching ratios $\brp(\pp \to \pz \hc) \times \brh(\hc \to \gamma \ec) =
[3.5 \pm 1.0~{\rm (stat)} \pm 0.7~{\rm (sys)}] \times 10^{-4}$.  When the $\ec$
is reconstructed in seven exclusive decay modes, $17.5 \pm 4.5$ $\hc$ events
are seen with an average mass $M(\hc) = [3523.6 \pm 0.9~{\rm (stat)}
\pm 0.5~{\rm (sys)}]$ MeV/$c^2$, and $\brp \brh = [5.3 \pm 1.5 {\rm~(stat)} \pm
1.0~{\rm (sys)}] \times 10^{-4}$.
If combined, the inclusive and exclusive data samples
yield an overall mass $M(\hc) = [3524.4 \pm 0.6~{\rm (stat)} \pm
0.4~{\rm (sys)}]$ MeV/$c^2$ and product of branching ratios $\brp \brh =
[4.0 \pm 0.8~{\rm (stat)} \pm 0.7~{\rm (sys)}] \times 10^{-4}$.  The $\hc$ mass
implies a P-wave hyperfine splitting $\Delta M_{\rm HF}(1P) \equiv \mav -
M(1^1P_1) = [1.0 \pm 0.6~(\stat) \pm 0.4~(\sys)]$ MeV/$c^2$.

\end{abstract}

\maketitle

Since the discovery of the $J/\psi$, the first bound state of a charmed quark
$c$ and charmed antiquark $\bar c$ \cite{Aubert:1974js,Augustin:1974xw}, the $c
\bar c$ ({\it charmonium}) spectrum has provided many insights about quarks and
the forces holding them together.  The charmed quark was the first to be found
with a mass larger than the characteristic scale of quantum chromodynamics
(QCD).  Charmonium bound states thus could be treated starting from a
nonrelativistic description \cite{Appelquist:1974zd}.  One could calculate
decay rates and level splittings and thereby determine the magnitude of the
strong coupling constant $\alpha_S$ at the charm mass scale, and the Lorentz
structure of the force confining quarks (see, e.g., \cite{Novikov:1977dq,%
Kwong:1987mj,Quigg:2004nv} for reviews.)

The hyperfine (spin-spin) splittings in charmonium S-wave states are
appreciable \cite{Eidelman:2004wy,etacmass}:
$$
\Delta M_{\rm HF}(1S) \equiv M(\jp) - M(\ec) \simeq 115~{\rm MeV}/c^2,
$$
\beq
\Delta M_{\rm HF}(2S) \equiv M(\pp) - M(\ec') \simeq 49~{\rm MeV}/c^2.
\eeq

For an interquark potential $V(r) = V_S(r)+V_V(r)$, the sum of vector $V_V(r)$
and scalar $V_S(r)$ contributions, only the vector part contributes to the
spin-spin splitting \cite{Novikov:1977dq,Kwong:1987mj,Ng:1985uq,Pantaleone:%
1985uf}, giving rise in lowest order of $1/m_c$ ($m_c$ is the mass of the
charmed quark) to a spin-spin interaction perturbation
\beq
V_{\rm SS}({\bf r}) =\frac{{\bf \sigma}_1 \cdot {\bf \sigma}_2}
{6 m_c^2} \nabla^2 V_V(r) = \frac{8 \pi \alpha_S {\bf \sigma}_1 \cdot
{\bf \sigma}_2}{9 m_c^2} \delta^3({\bf r}).
\eeq
The second equality on the right-hand side is obtained when one takes $V_V(r) =
4 \alpha_S/(3r)$ and neglects the slow variation of $\alpha_S$ with scale.  The
resulting local spin-spin interaction then contributes only to splittings in
S-wave states.  Taking account of the scale dependence of $\alpha_S$
\cite{Ng:1985uq,Pantaleone:1985uf} and $\chi_{cJ}$ wave function variations, one
finds at most a few MeV/$c^2$ splitting between the $1^1P_1$ state $\hc$ and
the spin-weighted average $\mav$ of the $^3P_J$ states $\chi_{cJ}$
\cite{Eidelman:2004wy}:
$\mav =  [M(1^3P_0) + 3 M(1^3P_1) + 5 M(1^3P_2)]/9 = (3525.4 \pm
0.1)$ MeV$/c^2$.  Small splittings $\Delta M_{\rm HF}(1P) \equiv \mav
- M(1^1P_1)$ are also consistent with a wide variety of estimates in
potential models \cite{models} and non-relativistic QCD \cite{NRQCD}, as well
as with lattice gauge theory estimates \cite{latt}.  Values of $|\Delta M_{\rm
HF}(1P)|$ larger than a few MeV/$c^2$ could indicate unexpected behavior of the
vector potential $V_V(r)$, unexpectedly large distortions of the masses of the
$1^3P_J = \chi_{cJ}$ states due to coupled-channel effects, or -- in lattice
theory -- effects of light-quark degrees of freedom. 

The low-lying charmonium spectrum is illustrated in Fig.\ 1.  The $\chi_{cJ}$
can be easily populated by radiative transitions from the $\pp$.
Their subsequent radiative decays to $\jp$ also are prominent.  In contrast,
the $h_c = 1^1P_1$ $c \bar c$ state is not easily produced.  It can be produced
in the $\bar p p$ direct channel, and a few events were seen at
the CERN Intersecting Storage Rings (ISR), clustered about $M(\hc) = 3525.4 \pm
0.8$ MeV/$c^2$~\cite{Baglin:1986yd}.  The significance of the signal was $2.3
\sigma$.  Stronger evidence was presented by Fermilab Experiment~E760 in
the channel $\bar p p \to h_c \to \pi^0 J/\psi$~\cite{Armstrong:1992ae}, with a
combined branching ratio
\beq
(1.7 \pm 0.4) \times 10^{-7} \le {\cal B}(h_c \to \bar p p) {\cal B}(h_c
\to \pi^0 J/\psi) \le (2.3 \pm 0.6) \times 10^{-7}
\eeq
for $M(h_c) = 3526.2 \pm 0.15 \pm 0.2$ MeV/$c^2$, with an additional possible
shift of up to $\pm 0.4$ MeV/$c^2$ due to resonance-continuum interference.
However, E835, the sequel to E760 with three times its integrated luminosity,
did not confirm the E760 signal \cite{Joffe,Andreotti:2005vu}.  Instead, a
signal with $\sim 3 \sigma$ significance for $\bar p p \to h_c \to \gamma
\eta_c \to \gamma \gamma \gamma$ was reported recently
\cite{Andreotti:2005vu}, with $M(h_c) = 3525.8 \pm 0.2 \pm 0.2$ MeV/$c^2$,
width $\Gamma \le 1$ MeV, and $(10.0 \pm 3.5) {\rm~eV} < \Gamma(h_c \to p \bar
p) {\cal B}(h_c \to \eta_c \gamma) < (12.0 \pm 4.5)$ eV.

\begin{center}
\begin{figure}
\includegraphics*[height=4.5in]{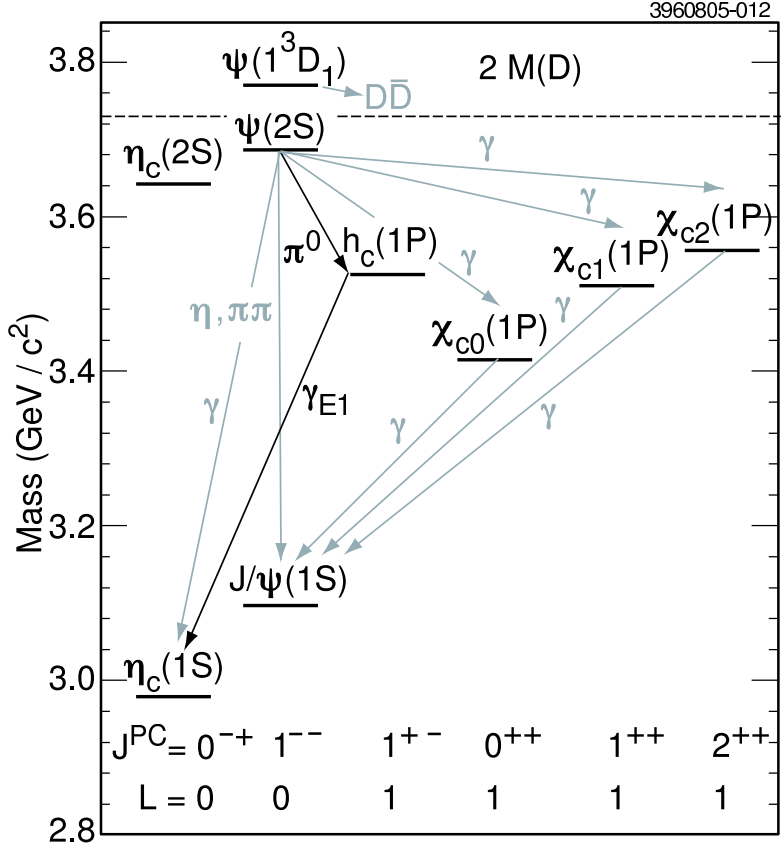}
\caption{The low-lying charmonium ($c \bar c$) spectrum and some observed
transitions.  The bold-faced lines labeled ``$\pz$'' and ``$\gme1$'' denote the
respective transitions $\pp \to \pz \hc$ and $\hc \to \gme1 \ec$ discussed in
the present paper.
\label{fig:charmon}}
\end{figure}
\end{center}

The decay $\pp \to \pz \hc$ can occur via isospin mixing (e.g., $\pz$--$\eta$
mixing) in the neutral pion \cite{Segre:1976wj}.  Previous experimental upper
limits on the branching ratio for this process are $\brp \equiv \br(\pp \to \pz
\hc) < 42$--80 $\times 10^{-4}$ for $M(\hc)$ between 3500 and 3535 MeV/$c^2$,
and $\brp \brh \ls 15 \times 10^{-4}$ for $M(\hc) \simeq 3525$ MeV/$c^2$,
where $\brh \equiv {\cal B} (\hc \to \gamma \ec)$ \cite{Porter:1982}.  Ko
\cite{Ko:1994nw} estimated $\brp \simeq 30 \times 10^{-4}$.  A recent
theoretical range is $\brp \simeq (4$--$13)\times 10^{-4}$ \cite{Kuang:2002hz}.

The decay $\hc \to \gamma \ec$ is an electric dipole (E1) transition whose
matrix element should be the same as that for the decays $\chi_{cJ} \to \gamma
\jp$.  Estimates \cite{E1rate}
of $\Gamma(\hc \to \gme1 \ec)$ range between 160 and 560 keV; a recent value
is 354 keV \cite{Godfrey:2002rp}.  The hadronic and photon + hadronic decay
rates of $\hc$ are not as well estimated, but the total width $\Gamma(\hc)$ is
generally found to be 1 MeV or less, with Ref.\ \cite{Godfrey:2002rp}
obtaining 0.94 MeV and hence $\brh \equiv \br(\hc \to \gme1 \ec) = 37.7\%$.
In other treatments this branching ratio can be larger; it is rarely smaller.
In  $\pp \to \pz \hc$ the polarizations of the $\hc$ and $\pp$ should be
almost identical, since the spinless $\pz$ is expected to be emitted in an
S wave.  The subsequent E1 transition $\hc \to \gamma \ec$ should then lead
to a photon with distribution $W(\cos \theta) \sim 1 + \cos^2 \theta$
with respect to the beam axis.

The present paper describes the identification of $\hc$ at the Cornell Electron
Storage Ring (CESR), using the CLEO III and CLEO-c detectors, via the
sequential process
\beq \label{eqn:proc}
e^+ e^- \to \pp(3686) \to \pz \hc~~,~~\hc \to \gme1 \ec~~,~~
\pz \to \gamma \gamma,
\eeq
illustrated by the bold arrows in Fig.\
\ref{fig:charmon} labeled ``$\pz$'' and ``$\gme1$,'' respectively.  Exclusive
reconstruction of $\ec$ decays in seven modes permits observation of $\hc$
with convincing significance and little background, while
inclusive analysis in which the $\ec$ is not reconstructed provides a better
measurement of $M(\hc)$ and of the combined branching ratio for $\pp \to \pz
\hc,~h_c \to \gamma \eta_c$.

We mention relevant aspects of the CLEO detector in Section II.  An overview of
inclusive and exclusive analysis methods is presented in Section III.  We then
describe background sources and suppressions (Sec.\ IV), data sample and event
selection (Sec.\ V), Monte Carlo samples (Sec.\ VI), the extraction of signal
from the data (Sec.\ VII), and systematic errors (Sec.\ VIII).  The combined
results of the different analyses are presented in Sec.\ IX.  A summary
and discussion of the results are given in Sec.\ X.
\bigskip

\centerline{\bf II.  THE CLEO DETECTOR}
\bigskip

The data upon which the present report is based were taken with the CLEO III
and CLEO-c detectors, described in detail elsewhere
\cite{Kubota:1991ww,Viehhauser:2001ue, Peterson:2002sk,Artuso:2002ya}.
Elements critical for the analyses presented here are the calorimeter and,
for the exclusive analysis, the charged particle
tracking and particle identification systems.  The barrel (80\% of $4 \pi$) and
endcap (additional 13\% of $4 \pi$) electromagnetic calorimeters consist of a
total of 7800 thallium-doped cesium iodide (CsI) crystals.  Their excellent
resolutions in position and energy (2.2\% at $E_\gamma=1$~GeV and 5\%
at 100~MeV) are a major source of sensitivity and discrimination against
background in identifying the chain of decays $\pp \to \pz \hc \to \pz
\gamma \ec$, and in measuring $M(\hc)$.  Pion/kaon separation is performed
utilizing the energy loss in the drift chamber, $dE/dx$, and photons in the
Ring-Imaging Cherenkov (RICH) counters. The combined $dE/dx$ and RICH particle
identification system has an efficiency of $>90\%$ and misidentification rates
of $<5\%$ for both $\pi^\pm$ and $K^\pm$.  Approximately one-half of the data
sample used an upgraded configuration, denoted CLEO-c, with an inner drift
chamber detector sensitive to longitudinal position \cite{Briere:2001rn}.
\bigskip

\centerline{\bf III.  OVERVIEW OF ANALYSES}
\bigskip

In the analyses described here one starts by looking for the
neutral pion emitted in $\pp \to \pz \hc$, expected to have an energy of
$E(\pz) \simeq 160$ MeV for $M(\pp) = 3686.111\pm0.025\pm0.009$ MeV/$c^2$
\cite{Aulchenko:2003qq} when $M(\hc) \simeq 3525$ MeV/$c^2$, and the
E1 photon emitted in $\hc \to \gme1 \ec$, with an expected energy in the $\hc$
rest frame of $\ege1 \simeq 502$ MeV for $M(\ec) = 2981.8 \pm 2.0$ MeV/$c^2$
\cite{etac}.  One takes advantage of the good energy resolution of
the CLEO electromagnetic calorimeter by searching for an enhancement in the
spectrum of masses $M(\hc)$ recoiling against the $\pz$,
\beq \label{eqn:mhc}
M(\hc) = [M^2(\pp) - 2 M(\pp) E(\pz) + M^2(\pz)]^{1/2},
\eeq
reducing background by selecting a range of E1 photon energy $\ege1$ or
$\eta_c$ mass $M(\ec)$ in the transition $\hc \to \gamma \ec$, with
\beq \label{eqn:mecr}
M(\ec) = \left \{ M^2(\hc) - 2 \ege1 [E(\hc) + p(\pz) \cos
\theta(\pz,\gme1)] \right \}^{1/2}.
\eeq
Here $E(\hc)$ and $p(\pz)$ are the $\hc$ energy and the magnitude of the
$\pz$ three-momentum in the $\pp$ rest frame, while $\theta(\pz,\gme1)$ is the
angle between the $\pz$ and $\gme1$ in that frame.

A search that is inclusive with respect to the $\eta_c$~decay, {\it i.e.,}~one
that imposes no further requirements on the $\eta_c$~decay products, exploits
the full event yield.  With a sample of approximately three million $\pp$, an
estimated product branching ratio $\brp \brh \simeq 4 \times 10^{-4}$, and an
estimated efficiency of about 15\%, one expects about 180 counts in the $\hc$
peak in inclusive analyses, albeit on top of a background several times larger.

An exclusive analysis, in which specific decay modes of the $\eta_c$ are
detected, benefits from much lower backgrounds with reduced efficiency.  In the
present analysis nearly 10\% of all $\eta_c$ decays are reconstructed, leading
one to expect $\sim 18$ events with little background.  The method is validated
by reconstructing the more abundant $\eta_c$ decays in the direct reaction
$\pp \to \gamma \ec$, for which $\br(\pp \to \gamma \ec) = (3.2 \pm 0.6 \pm
0.4) \times 10^{-3}$ \cite{Athar:2004dn}.  (The Particle Data Group average of
other measurements is $(2.8 \pm 0.6) \times 10^{-3}$ \cite{Eidelman:2004wy}.)

The following features are common to both inclusive and exclusive analyses.
The sensitivity of the search for $\pp \to \pi^0 h_c \to \pi^0 \gamma \eta_c$
depends upon the degree to which the $\pi^0$ peak can be recognized above a
background which rises sharply as $\pi^0$ energy increases.  Thus understanding
of $E(\pi^0)$ resolution is central to observation of the $h_c$ in this
process.  It is also crucial in pinning down the mass of $h_c$. 

Because the signal $\pi^0$ in $\pp \to \pi^0 h_c$ is expected to have fairly
low momentum, its decay photons tend to be back-to-back in azimuth.
Mismeasurements of their energies are partly compensated by the mass constraint
used when combining them into a $\pi^0$~candidate and thus affect the $\pi^0$
detection probability only minimally, resulting in a narrow distribution in
$\pz$ energy and therefore in $M(h_c)$, as will be seen in the specific
analyses described below. 

At $E_\gamma \simeq 500$ MeV (the energy of the expected signal for $\hc \to
\gme1 \ec$), the experimental resolution of the photon energy is comparable
to that expected from Doppler broadening of the $h_c$ when the photon is
observed in the $\pp$~rest frame ($\sim 10$~MeV).  One can correct for this
broadening using information on $\cos \theta(\pz,\gme1)$ as in Eq.\
(\ref{eqn:mecr}).

Two complementary inclusive analyses have been pursued.  In one, candidates for
$\cascade$ are selected by choosing events containing an E1 photon candidate in
a range of energies expected for $\hc \to \gme1 \ec$, and displaying a peak in
$M(\hc)$.  This method has the advantage that backgrounds to the signal photon
and $\pz$ are uncorrelated with one another, but it presupposes foreknowledge
of the interesting range of $M(\hc)$ values, and does not compensate for the
broadening of the photon energy spectrum due to $\hc$ recoil.
In a second inclusive method, events are chosen within a given range of
$M(\ec)$ as calculated from the energies and relative angle of the $\pz$ and
$\gme1$, and displays a peak in $M(\hc)$.  This method compensates for the
recoil broadening of the $\gme1$ energy spectrum and does not presuppose a
value of $M(\hc)$.  However, since both photon and $\pz$ energies are needed to
calculate $M(\ec)$, backgrounds are correlated, and some subtraction methods
appropriate for the first method are not valid for the second.

Exclusive reconstruction of decay modes of the $\ec$ offers the potential of
significant background reduction.  The following $\eta_c$~decay modes were
studied: 
$K_S^0 K^\pm \pi^\mp$,
$K_L^0 K^\pm \pi^\mp$,
$K^+ K^- \pi^+ \pi^-$,
$\pi^+ \pi^- \pi^+ \pi^-$,
$K^+ K^- \pi^0$,
$\pi^+ \pi^- \eta(\gamma\gamma)$,
$\pi^+ \pi^- \eta(\pi^+\pi^-\pi^0)$.
They are summarized in Table \ref{tab:ecmodes} together with their branching
fractions in $\eta_c$ decay \cite{Eidelman:2004wy}.  In order to reduce the
effect of the poorly known $\ec$ branching ratios, the ratio of rates of $\pp$
decay to $\pi^0 \gamma \eta_{c}$ and $\gamma \eta_{c}$ is measured. The
normalizing mode has been recently measured at CLEO \cite{Athar:2004dn}.  Its
study also permits us to construct and verify event selection criteria in $\ec$
reconstruction.

\begin{table}[h]
\caption{Decay modes of $\ec$ used in the exclusive analysis and their 
branching fractions ${\cal B}$~\cite{Eidelman:2004wy}.
\label{tab:ecmodes}}
\begin{center}
\begin{tabular}{l c} \hline \hline
Mode                                & ${\cal B}$ (\%)     \\ \hline
$\ks K^\pm \pi^\mp$                 & $1.9 \pm 0.5$       \\
$\kl K^\pm \pi^\mp$                 & $1.9 \pm 0.5$       \\
$K^+ K^- \pi^+ \pi^-$               & $1.5 \pm 0.6$       \\
$\pi^+ \pi^- \pi^+ \pi^-$           & $1.2 \pm 0.3$       \\
$K^+ K^- \pi^0$                     & $1.0 \pm 0.3$       \\
$\pi^+ \pi^- \eta(\gamma \gamma)$   & $1.3 \pm 0.5$       \\
$\pi^+ \pi^- \eta(\pi^+\pi^-\pi^0)$ & $0.7 \pm 0.3$       \\ \hline
Total                               & $9.5 \pm 1.6$       \\ 
\hline \hline
\end{tabular}
\end{center}
\end{table}

\bigskip
\centerline{\bf IV.  BACKGROUND SOURCES AND SUPPRESSIONS}
\bigskip

We first describe major backgrounds to the signal, and how they are suppressed,
in a qualitative manner.  Details of background suppression are described in
the next section. Selection criteria are applied in different ways depending on
the nature of the analysis.

\begin{itemize}

\item {\it The transition $\pp \to \pi^+ \pi^- \jp$.}
Approximately 1/3 of all $\pp$ decay to the final state $\pi^+ \pi^-
\jp$ \cite{Adam:2005uh}.  Subsequent decays of $\jp$ can generate both soft
$\pz$s (a background to the signal for $\pp \to \pz \hc$) and hard
photons in the vicinity of the signal energy $\ege1 \simeq 500$ MeV for the
expected E1 transition $\hc \to \gme1 \ec$.  Thus, all analyses to be reported
here excluded some range of mass $X$ around $M(\jp)$ recoiling against $\pi^+
\pi^-$ in the reaction $\pp \to \pi^+ \pi^- X$.

\item {\it The transition $\pp \to \pz \pz \jp$.}
The decay $\pp \to \pz \pz \jp$ accounts for about 1/6 of all $\pp$ decays
\cite{Adam:2005uh}.  In addition to the backgrounds mentioned above for charged
pion pairs, either of the two neutral pions can be mistaken for that in the
signal for $\pp \to \pz \hc$.  Thus, in inclusive analyses, a range of masses
around $M(\jp)$ in the spectrum recoiling against the dipion pair in
$\pp \to \pz \pz X$ was excluded.

\item {\it The transition $\pp \to \gamma \chi_{cJ} \to \gamma \gamma \jp$.}
The sum of the product branching ratios $\br(\pp \to \gamma \chi_{cJ})
\br(\chi_{cJ} \to \gamma \jp)$ exceeds 5\% \cite{Adam:2005uh}.  This
background can be reduced by excluding events with a range of masses 
around $M(J/\psi)$ in the spectrum recoiling against $\gamma \gamma$ in
$\gamma \gamma X$.

\item {\it Candidates for 500 MeV E1 photons which are $\pz$ or $\eta$
decay products.}
A sufficiently energetic $\pz$ can give rise to a photon which can
be mistaken for the signal E1 photon in $\hc \to \gme1 \ec$.  It is possible
to suppress such photons by rejecting all candidates which can form a
candidate $\pz$ if paired with another photon.  A similar rejection of $\eta$
decay products also can be applied.

\item {\it Mis-pairings of candidates for $\pz$ decay.}
In general photons from $\pz$ decays are identified by requiring that their
energies and directions lead to a reconstructed $\pz$ mass within about
15 MeV/$c^2$ of the nominal value of 135 MeV/$c^2$.
If some other pairing gives a better-reconstructed $\pz$ mass,
the original pairing is discarded and the better pairing is adopted.

\end{itemize}
\bigskip

\centerline{\bf V. DATA SAMPLE AND EVENT SELECTION}
\bigskip

The data samples obtained with the CLEO III and CLEO-c configurations are shown
in Table \ref{tab:data}, where the number of events was calculated by the method
described in \cite{Athar:2004dn} and was estimated to have an uncertainty of
$\pm 3\%$.

\begin{table}
\caption{Conditions under which $\pp$ data were acquired for this
analysis.  Here $\Delta E_{\rm cm}$ denotes the center-of-mass energy spread,
while $\int {\cal L} dt$ denotes integrated luminosity
measured using the reaction $e^+e^- \to \gamma\gamma$.
\label{tab:data}}
\begin{center}
\begin{tabular}{l c c c c} \hline \hline
Detector &  Time   & $\Delta E_{\rm cm}$ & $\int {\cal L} dt$ & $N(\pp)$ \\ 
         & period  &(MeV) &(pb)$^{-1}$        & $(10^6)$ \\ \hline
CLEO-III & 2002--3 & 1.5 &       2.74        &   1.56   \\
CLEO-c   & 2003--4 & 2.3 &       2.89        &   1.52   \\
Total    &         &         &       5.63        &   3.08   \\ \hline \hline
\end{tabular}
\end{center}
\end{table}
\bigskip

Common features of event selection for all analyses are listed in the
following.  Several other analysis-specific criteria will be described in the
corresponding subsections.  Selection requirements for all analyses
are summarized in Table \ref{tab:evtsel}.

\begin{itemize}

\item Charged particle selection criteria were standard ones used for other
CLEO analyses.  The distance of closest approach of a track with respect to the
run-averaged collision point was required to be less than 5 cm along the beam
line and less than 0.5 cm in the direction transverse to the beam.  Each track
was required to be fitted with a reduced $\chi^2$ (i.e., per degree of freedom)
of less than 20, to give between 50\% and 120\% of the expected number of
signals on drift chamber wires, and to make an angle of at least $21.6^\circ =
\cos^{-1} (0.93)$ with respect to the beam axis.

\item A photon candidate was defined as a shower
which does not match a track within
100 mrad, is not in a ``hot'' cell of the electromagnetic calorimeter, and has
the transverse distribution of energy consistent with an electromagnetic shower.

\item The minimum $\pi^0$ photon candidate energy was set at 30 MeV in the
barrel and 50 MeV in the endcaps.

\item In kinematic fitting, photon energies and angles for $\pi^0$ candidates
were adjusted to give the exact $\pi^0$ mass.  This increases precision in the
determination of the $\pz$ energy and hence the $h_c$ mass, which is computed
from Eq.\ (\ref{eqn:mhc}) using the nominal values of $M(\pp)$ and $M(\pz)$.

\item Photon candidates for the E1 transition $\hc \to \gamma \ec$ were
subjected to background suppression involving vetoing of candidates which could
form a $\pz$.

\item Neutral pion candidates were tested for the possibility that one of their
showers could form a neutral pion with some other shower, and were rejected if
any other pairing was more consistent with a $\pz$ mass.

\item Events were flagged if they were candidates for the
processes $\pp \to \pi^+ \pi^- \jp$ or $\pp \to \pi^0 \pi^0 \jp$
and rejected accordingly.

\item When an empirical parametrization of the background shape was needed,
the analyses employed a convenient parametrization of backgrounds to the $\pz$
recoil spectrum known as an ARGUS function \cite{ARGUSfn}, appropriate for
processes such as $\pp \to \pz \hc$ in which there is a kinematic endpoint, 
equal here to $M(\pp) - M(\pi^0) = 3551.2$ MeV/$c^2$.

\item A large generic Monte Carlo sample of $\simeq 39$ million $\pp$ events
permitted the optimization of signal-to-background ratio by adding an
appropriately normalized sample of signal Monte Carlo events and choosing
event selection criteria to maximize the likelihood ratio for fits with and
without a resonance signal.

\item  The distribution of the photon polar angles in both $h_c \to \gamma \ec$
and $\pp \to \gamma \ec$ (relevant to the exclusive analysis) was assumed to be
$\sim 1 + \cos^2 \theta$.  For the former decay this assumption is based on the
expectation that the $h_c$ retains the $\pp$ polarization in the
(mainly S-wave) process $\pp \to \pz \hc$.

\end{itemize}

\begin{table}
\caption{Comparison of event selection criteria for inclusive and exclusive
analyses.
\label{tab:evtsel}}
\begin{center}
\begin{tabular}{|l|c|c|c|} \hline \hline
Property & \multicolumn{2}{c|}{Inclusive analysis specifying:} & Exclusive \\
\cline{2-3}
or quantity & $\ege1$ range & $M(\ec)$ range & analysis \\ \hline
Initial & $\ge 2$ charged & Depends on \#         & Hadronic \\
event   & tracks and      & ($\ge 1)$ of charged & selection\\
selection & $\ge 3$ showers & tracks (see text)  & (see text)\\ \hline
$\ege1$ or &   $\ege1 = $   &   $M(\ec) \pm35$ & $M(\ec) \pm50 $ \\
$M(\ec)$range & $503\pm35$ MeV & MeV/$c^2$ & MeV/$c^2$ \\ \hline
Photon & 10 most & All & All \\
showers & energetic & & \\ \hline
Photon & Barrel plus & Barrel & Barrel plus \\
acceptance & endcaps & only & endcaps \\ \hline
No.\ of $\pz$ in  & One and  & One and  & At least \\
signal region (a) & only one & only one & one      \\ \hline
$\pz$ rejection & Reject best- & Reject all $\pz$ & Reject all $\pz$ \\
on $\gme1$      & pull $\pz$ only & with pull $\le 2.5$ & with pull $\le 3$
\\ \hline
$\eta$ rejection & $M(\gme1 \gamma) =$  & None & None \\
on $\gme1$      & $550\pm25$ MeV/$c^2$ & & \\ \hline
$|\Delta M(\pi^+\pi^-J/\psi)|$ & $\le 15$ & $\le 8.4$ & $\le 10$ \\
excluded & MeV/$c^2$ & MeV/$c^2$ & MeV/$c^2$ \\ \hline
$|\Delta M(\pz \pz J/\psi)|$ & $\le 40$ & $\le 32$ & None \\ 
excluded & MeV/$c^2$ & MeV/$c^2$ & \\ \hline
$\gamma \gamma J/\psi$ & $M({\rm all~chgd})$ &
 $|\Delta M(\gamma \gamma J/\psi)|$ & None \\
rejection & $\ge 3050$ MeV/$c^2$ & $\ge 40$ MeV/$c^2$ & \\ \hline \hline
\end{tabular}
\end{center}
\leftline{(a) Defined as giving $M(\hc) = 3526 \pm 30$ MeV/$c^2$}
\end{table}

\bigskip
\leftline{\bf A.  Inclusive analyses}
\bigskip

The event selection criteria for the analysis selecting a range of $\ege1$ are
summarized in Table \ref{tab:evtsel}.  Showers
were required to have at least 30 MeV energy if detected in the barrel region
of the calorimeter and at least 50 MeV if detected in the endcaps.  Only the
ten highest-energy showers and tracks in an event were considered, in order to
reduce combinatorial background.  A maximum of ten neutral pions composed of
the ten highest-energy showers was considered.

Neutral pions were reconstructed by requiring that the two-photon invariant
mass be in the range $M_{\gamma \gamma} = 135 \pm 15$ MeV/$c^2$ or
within three standard deviations of the peak.  (Resolutions in MeV/$c^2$
depend on properties of each candidate, such as energy and calorimeter
location.)

Selection criteria were guided by maximizing the likelihood ratio for fits to
Monte Carlo-generated background with and without a simulated signal.  In order
to reduce the abundant background due to photons and charged particles from the
decay of $J/\psi$, the cascades $\pp \to J/\psi X$ were suppressed by excluding
candidates for $\pp \to (\pi^+ \pi^- \jp,~\pz \pz \jp, \gamma \gamma \jp)$
using the criteria in the second column of Table \ref{tab:evtsel}.  Photon
candidates for $\gme1$ in $\hc \to \gme1 \ec$
were rejected if they could form a $\pz$ or $\eta$ (defined, respectively,
by $M_{\gamma\gamma} = 135 \pm 15$ or $550 \pm  25$ MeV/$c^2$) when combined
with any other photon.  It was demanded that there be only one photon in the
event with energy $503 \pm 35$ MeV.

In the complementary analysis selecting a range of $M(\ec)$ (Table
\ref{tab:evtsel}, third column), events were chosen corresponding to a slight
modification of a previously used criterion \cite{Athar:2004dn} for selection
of hadronic events at the $\pp$ energy.\footnote{For $1 \le N_{\rm ch} \le 3$ 
($N_{\rm ch}$ = number of charged tracks), the maximum energy visible in the
calorimeter was required to be less than the total center-of-mass energy
$E_{\rm CM}$, vs.\ $0.85E_{\rm CM}$ in Ref.\ \cite{Athar:2004dn}.  For
$N_{\rm ch} \ge 4$ the criteria were the same as in Ref.\ \cite{Athar:2004dn}.}
Background suppression techniques were similar in most respects to those of
the other inclusive analysis except for the following details:

\begin{itemize}

\item Photons for $\pi^0$ or $\gme1$ candidates were chosen only in the barrel
region of the electromagnetic calorimeter, in an attempt to improve energy
resolution.

\item Neutral pion candidates were required to have a $\gamma \gamma$ mass
within $2.5 \sigma$ of the peak, and were rejected if any other pairing of
photons within this same ``pull mass'' (normalized deviation from the correct
mass in units of Gaussian width) provided a better fit to the $\pi^0$ mass. 
Partner photons for this rejection were allowed to be either in endcaps ($E >
50$ MeV) or barrel ($E > 30$ MeV).

\item Candidates for the E1 transition photon which could form a $\pi^0$ were
vetoed \cite{Athar:2004dn} as in the $\ege1$-selection analysis, rejecting
any photon forming a pair with mass less than $2.5 \sigma$ from $M(\pz)$
when combined with a photon in endcap regions of the calorimeter with
at least 50 MeV or barrel regions with at least 30 MeV.  However, Monte Carlo
simulations (to be discussed in Sec.\ VI) indicated no need to veto $\eta$
mesons.

\end{itemize}
\newpage

\leftline{\bf B.  Exclusive analysis}
\bigskip

The exclusive analysis measures the ratio of the cascade decays $\cascade$ to
the direct radiative decays $\direct$ by identifying the decay channels listed
in Table I.  To design event selection criteria, 20,000 signal
Monte Carlo were generated for each mode of the cascade and direct radiative
decays.  The 39 million generic Monte Carlo $\pp$ decays without $h_{c}$
were utilized to study the background to the cascade decay.  All reconstructed
events were required to have no extra tracks and total extra unmatched shower
energy less than 200~MeV.  The basic particle selection criteria,
in addition to those mentioned at the start of this Section, include the
following specific to this analysis:

\begin{itemize}

\item $\pi^{0}$: Mass less than $3 \sigma$ from nominal value.

\item $\ks$: Decay displaced by more than $3 \sigma$ with respect to the
run-averaged collision
point, mass within 10~MeV/$c^2$ of nominal value

\item $\eta(\gamma\gamma)$: Mass within $3 \sigma$ of the nominal $\eta$ value 

\item $\eta(\pi^{+}\pi^{-}\pi^{0})$ : $M_{\pi^{+}\pi^{-}\pi^{0}}$ within 20
MeV/$c^2$ of the nominal $\eta$ mass

\end{itemize}

Information from the RICH and $dE/dx$ detectors was combined to distinguish
kaons from pions when RICH information was available.  RICH information was
utilized when a track was in the RICH fiducial volume with $|\cos\theta| <
0.8$, a kaon candidate had momentum at least 600 $\MeV/c$, and three or more
photons were detected near the predicted ring location.  A combined
``Log-Likelihood'' was defined as
$\Delta L = L(\pi)_{\rm RICH}-L(K)_{\rm RICH} + (\sigma_{dE/dx}^{\pi})^{2} -
(\sigma_{dE/dx}^{K})^{2}$, where $L(\pi)_{\rm RICH}$ is $-2$ times
the natural logarithm of the RICH likelihood
for the pion hypothesis, and $L(K)_{\rm RICH}$ is for the kaon hypothesis,
while $\sigma_{dE/dx}^{\pi}$ is the deviation of $dE/dx$ from what is expected
for the pion hypothesis normalized to the measurement error and
$\sigma_{dE/dx}^{K}$ is the same for the kaon hypothesis.  If RICH information
was not available, a track was identified as a kaon if $|\sigma_{dE/dx}^{K}| <
3$ and $|\sigma_{dE/dx}^{K}| <|\sigma_{dE/dx}^{\pi}|)$.  When RICH information
was not available and track momentum was above 600 $\MeV/c$, a track was
identified as a pion if $|\sigma_{dE/dx}^{\pi}| < 3$.  When RICH information
was available or track momentum was below 600 $\MeV/c$, charged kaons and pions
were well-separated. In the $\kkpp$ and $\kkpizero$ modes, at least one kaon
candidate was required to be identified when $K$ and $\pi$ were well-separated.

Because the $\pp$ resonance width is only 0.3 MeV, considerably less than the
beam energy spread, the beam energy was always assumed to be half of $M(\pp)$
when running at the $\pp$.\footnote{The crossing angle is around 4 mrad,
corresponding to a transverse momentum of about 3686 $\sin(0.004) =
15$ MeV/$c$.} In $\ec\to \klkp$, the missing mass should equal the $\kl$
nominal mass since the $\kl$ is undetected. In this case, a 1C kinematic fit
was performed assuming that the missing particle has the mass of $\kl$.  In all
other modes, $\pp$ final decay particles were fully
reconstructed, and the net 4-momentum of reconstructed charged or neutral
tracks should equal the 4-momentum of the $\pp$ which is known, permitting
4C kinematic fits. The $\chi^{2}$ values from the fits indicate how well each
reconstructed event matches the kinematics of the decay hypothesis.  A rather
loose requirement of $\chi^{2}$/d.o.f.$<10$ in all modes was imposed.
The $\ec$ signal was fully reconstructed in all the modes except $\klkp$. In
$\klkp$, the $\eta_{c}$ mass was inferred from the energies of the recoiling E1
photon and $\pz$.

Generic Monte Carlo studies indicate that photons from $\pz$s in $\pp \to \pz
\pz J/\psi$ and $ \pp \to \gamma \chi_{cJ}$ ($\chi_{cJ} \to \pz X$) decays are
a large background source to $\gme1$.  A photon candidate was vetoed if the
absolute value of its best $\pz$ pull mass, when combined with all other
photons of energies greater than 30 MeV, was less than 3. This cut greatly
reduced the background but also resulted in a 15\% efficiency loss according to
signal Monte Carlo.  The net effect on the expected sensitivity to $\hc$ was
positive.
\bigskip

\centerline{\bf VI.  MONTE CARLO SAMPLES}
\bigskip

Monte Carlo simulations of background and signal were employed in order to
optimize event selection criteria and to estimate backgrounds to data.
The generic Monte Carlo sample mentioned earlier was used.  Simulations
employed hadronization routines embodied in JETSET \cite{JETSET}, with its
parameters optimized for $\pp$ decays \cite{Athar:2004dn}.  The detector
simulation was based on Geant \cite{GEANT}.  Hadronization of $h_c$ decays
was emulated using Model 14 of the LUND/JETSET fragmentation algorithm.
\bigskip

\leftline{\bf A.  Inclusive analyses}
\bigskip

\leftline{\it 1.  Choice of background shapes.}

The $\ege1$-range analysis uses the $\pz$ recoil spectrum from the data itself
as background, without demanding a candidate with $E_\gamma = 503 \pm 35$ MeV
for the E1 photon.  This is feasible since the $h_c$ contribution is invisible,
being at the level of $\sim 4 \times 10^{-4}$.  The $M(\ec)$-range analysis
uses generic Monte Carlo background instead, since the selection of an $\ec$
mass range in analyzing the data affects the background shape.
\bigskip

\leftline{\it 2.  Optimization of signal significance.}

Monte Carlo samples were employed to choose ranges of selection
providing the highest sensitivity to the $\hc$ signal, as judged by
maximum likelihood for the resonance hypothesis.   These samples also
permitted studies of input/output agreement and statistical variation.
The optimum event selection criteria determined in these Monte Carlo studies
were applied to the data.

In the $\ege1$-range analysis, 30,000 signal events were generated for
$\cascade$.  Assuming $\brp \brh \equiv \br(\pp \to\pi^0h_c) \times
\br(h_c\to\gamma\eta_c) = 4.0 \times 10^{-4}$, 15,600 signal events were added
to the 39 million generic Monte Carlo
sample. The input masses and widths were taken as $M(h_c) = 3526$ MeV/$c^2$,
$\Gamma (h_c) = (0.5,~0.9,~1.5) $ MeV, and $M(\ec)=2982$ MeV/$c^2$, $\Gamma
(\ec)=24.8$ MeV \cite{etac}.  In the $M(\ec)$-range analysis, $185 \times 10^3$
events were generated for $\pp \to \pi^0 \hc$, with a 37.7\% branching ratio
\cite{Godfrey:2002rp} for the subsequent decay $\hc \to \gamma \ec$.  The
remaining $\hc$ decays were taken to have a 56.8\% branching ratio to $ggg$
and a 5.5\% branching ratio to $\gamma gg$.  The mass of $h_c$ was assumed to
be 3525.3 MeV/$c^2$, and the $h_c$ width was taken to be 1 MeV.  The mass of
$\eta_c$ was chosen as 2981.8 MeV/$c^2$ \cite{etac}.

The results of the Monte Carlo studies for the $\ege1$-range analysis are
summarized in the second and third columns of Table \ref{tab:summ}.
Significance levels are obtained as $\sigma\equiv\sqrt{-2\ln(L_0/L_{\rm max})}$,
where $L_{\rm max}$ is the maximum likelihood for the resonance fit, and $L_0$
is the likelihood for the fit with no $h_c$ resonance.  Selection ranges
(summarized in the second column of Table \ref{tab:evtsel}) were chosen to
maximize the significance for the Monte Carlo sample calculated in this manner.
For each effect examined, asterisked values for all other parameters were
assumed.

\begin{table}
\caption{Results of Monte Carlo optimizations using a combined sample of
39 million generic $\pp$ events and 15,600 signal events for $\ege1$-range
analysis.  Asterisks show final selection.
\label{tab:summ}}
\begin{center}
\begin{tabular}{|c|c|c|c|c|c|c|c|}\hline \hline
 & \multicolumn{2}{c|}{MC} & \multicolumn{5}{c|}{DATA}\\ \hline
 & Signif. $(\sigma)$ & $s^{2}/B$ & Mass, MeV/$c^2$ & Yield & $\brp
 \brh \times 10^{4}$ & $\chi^{2}$/DOF & Signif. $(\sigma)$  \\ \hline \hline

\multicolumn{8}{|c|}{Effect of background shapes} \\ \hline
$\ast$ DATA & & &  3524.4$\pm$0.7 &  139$\pm$41 &  3.4$\pm$1.0 & 
 1.36 &  3.6 \\
  MC  & & &  3524.6$\pm$0.7 &  146$\pm$40 &  3.5$\pm$1.0 & 
 1.59 &  3.8  \\ \hline

\multicolumn{8}{|c|}{All of the following optimizations were done using
 background from DATA} \\ \hline

\multicolumn{8}{|c|}{Effect of changing range of hard $\gamma$ energy,
 503$\pm$, MeV/$c^2$} \\ \hline
$\pm$30 & 16.4 & 1.01 & 3524.0$\pm$0.7 & 120$\pm$38 & 3.1$\pm$0.9 &
 1.19 & 3.3  \\
$\ast$ $\pm$35 & 17.3 & 1.00 & 3524.4$\pm$0.7 & 139$\pm$41 & 3.4$\pm$1.0 &
 1.36 & 3.6  \\
$\pm$40 & 16.1 & 0.96 & 3524.4$\pm$0.6 & 145$\pm$43 & 3.4$\pm$1.0 &
 1.28 & 3.5 \\
$\pm$45 & 16.3 & 0.90 & 3524.8$\pm$0.8 & 134$\pm$45 & 3.0$\pm$1.0 &
 1.24 & 3.1 \\
$\pm$50 & 15.8 & 0.86 & 3524.8$\pm$0.9 & 132$\pm$47 & 2.9$\pm$1.0 &
 1.26 & 2.9 \\ \hline

\multicolumn{8}{|c|}{Effect of changing mass range for $\pi^+\pi^- J/\psi$
rejection, MeV/$c^2$} \\ \hline
$\pm6$  & 17.2 & 0.97 & $3524.4\pm0.6$ & $158\pm43$ & $3.7\pm1.0$ &
 1.28 & 3.9 \\
$\pm10$ & 17.3 & 0.99 & $3524.3\pm0.6$ & $156\pm42$ & $3.7\pm1.0$ &
 1.36 & 3.9 \\
$\ast$ $\pm15$ & 17.3 & 1.00 & $3524.4\pm0.7$ & $139\pm41$ & $3.4\pm1.0$ &
 1.36 & 3.6  \\
$\pm20$ & 17.1 & 1.00 & $3524.2\pm0.7$ & $132\pm40$ & $3.3\pm1.0$ &
 1.38 & 3.4 \\ \hline

\multicolumn{8}{|c|}{Effect of changing mass range for $\pi^0\pi^0 J/\psi$
rejection, MeV/$c^2$} \\ \hline
$\pm20$ & 17.2 & 0.99 & $3524.3\pm0.8$ & $140\pm42$ & $3.3\pm1.0$ &
 1.45 & 3.4 \\
$\pm30$ & 17.2 & 1.00 & $3524.5\pm0.8$ & $134\pm41$ & $3.2\pm1.0$ &
 1.30 & 3.4 \\
$\ast$ $\pm40$ & 17.3 & 1.00 & $3524.4\pm0.7$ & $139\pm41$ & $3.4\pm1.0$ &
 1.36 & 3.6  \\
$\pm50$ & 17.3 & 1.00 & $3524.4\pm0.7$ & $147\pm41$ & $3.6\pm1.0$ &
 1.30 & 3.8 \\ \hline

\multicolumn{8}{|c|}{Effect of number of $\pi^0$s in the signal region}
 \\ \hline
$\ast$ $=1$ & 17.3 & 1.00 & $3524.4\pm0.7$ & $139\pm41$ & $3.4\pm1.0$ &
 1.36 & 3.6  \\
$\ge 1$ & 17.2 & 0.95 & $3524.8\pm0.9$ & $122\pm42$ & $2.9\pm1.0$ &
 1.04 & 3.0 \\ \hline

\multicolumn{8}{|c|}{Effect of endcap $\gamma$s in signal $\pi^0$s} \\ \hline
$\ast$ with & 17.3 & 1.00 & $3524.4\pm0.7$ & $139\pm41$ & $3.4\pm1.0$ &
 1.36 & 3.6  \\
without & 16.0 & 0.91 & $3524.8\pm0.7$ & $123\pm37$ & $3.4\pm1.0$ &
 1.16 & 3.5 \\ \hline

\multicolumn{8}{|c|}{Effect of $\eta$ suppression on E1 photon} \\ \hline
$\ast$ with & 17.3 & 1.00 & $3524.4\pm0.7$ & $139\pm41$ & $3.4\pm1.0$ &
 1.36 & 3.6  \\
without     & 15.8 & 0.91 & $3524.6\pm0.8$ & $135\pm45$ & $3.0\pm1.0$ &
 1.21 & 3.1 \\ \hline

\multicolumn{8}{|c|}{Effect of $\pp \to \gamma \chi_{1,2} \to
 \gamma\gamma J/\psi$ suppression } \\ \hline
$\ast$ without & 17.3 & 1.00 & $3524.4\pm0.7$ & $141\pm41$ & $3.4\pm1.0$ &
 1.36  & 3.6  \\
with     & 17.0 & 1.02 & $3524.6\pm0.7$ & $137\pm40$ & $3.4\pm1.0$ &
 1.21 & 3.6 \\ \hline \hline
\end{tabular}
\end{center}
\end{table}

\begin{center}
\begin{tabular}{|c|c|c|c|c|c|} \hline \hline
 \multicolumn{6}{|c|}{Table IV, continued} \\ \hline
 & \multicolumn{5}{c|}{DATA (no MC entries)}\\ \hline
 & Mass, MeV/$c^2$ & yield & $\brp \brh \times 10^{4}$
 & $\chi^{2}$/DOF & signif. $(\sigma)$  \\ \hline \hline

\multicolumn{6}{|c|}{Effect of changing total width of $h_c$, MeV} \\ \hline
$0.5$ & $3524.3\pm0.7$ & $132\pm38$ & $3.2\pm0.9$ & 1.36 & 3.6 \\
$\ast$ 0.9 & $3524.4\pm0.7$ & $139\pm41$ & $3.4\pm1.0$ & 1.36 & 3.6  \\
$1.5$ & $3524.5\pm0.7$ & $149\pm44$ & $3.6\pm1.1$ & 1.39 & 3.6 \\ \hline

\multicolumn{6}{|c|}{Effect of changing $\pi^0$ resolution widths} \\ \hline
$\ast$ MC &  $3524.4\pm0.7$ & $139\pm41$ & $3.4\pm1.0$ & 1.36 & 3.6 \\
MC-25\% & $3524.3\pm0.6$ & $131\pm38$ & $3.2\pm0.9$ & 1.35 & 3.6 \\
MC+25\% & $3524.5\pm0.7$ & $149\pm45$ & $3.6\pm1.1$ & 1.39 & 3.6 \\ \hline

\multicolumn{6}{|c|}{Effect of binning} \\ \hline
$\ast$ 2 MeV/$c^2$ & $3524.4\pm0.7$ & $139\pm41$ & $3.4\pm1.0$ & 1.36 & 3.6 \\
1 MeV/$c^2$ & $3524.5\pm0.6$ & $137\pm41$ & $3.3\pm1.0$ & 1.16 & 3.5 \\ \hline 

\multicolumn{6}{|c|}{Effect of changing fit range, MeV/$c^2$} \\ \hline
$\ast$ 3496-3552 & $3524.4\pm0.7$ &  $139\pm41$ &  $3.4\pm1.0$ & 1.36 & 3.6 \\
3500-3540 & $3524.4\pm0.7$ & $139\pm42$ & $3.4\pm1.0$ & 0.96 & 3.5 \\ \hline

\multicolumn{6}{|c|}{CLEO III VERSUS CLEO-c} \\ \hline
CLEO III & $3523.8\pm0.7$ & $94\pm30$ & $4.5\pm1.4$ & 0.96 & 3.3 \\
CLEO-c & $3526.1\pm1.5$ & $56\pm28$ & $2.8\pm1.4$ & 1.55 & 2.1 \\
\hline \hline
\end{tabular}
\end{center}
\bigskip
\bigskip

\noindent
These choices were found to lead to the same output from the $\pp$ generic
Monte Carlo sample as the input:

\begin{center}
\begin{tabular}{|c|c|c|} \hline \hline
 & Input & Output \\ \hline
$M(h_{c})$ (MeV/$c^2$) & 3526.0 & $3525.9\pm0.1$ \\
$\brp \brh \times 10^4$ & 4.0 & $4.1 \pm 0.3$ \\ \hline \hline
\end{tabular}
\end{center}

The above choices were based on maximum likelihood in 22 variations with no
contact with the experimental data, i.e., by ``blind'' analysis.  The best
choices indeed are mirrored in the data.  Table \ref{tab:summ} therefore
lists for the data the values of the likelihood-based significance for
all 22 variations examined in the Monte Carlo sample.  It is interesting to
note that these choices do lead to higher significance values in most cases,
although, as is to be expected, because of the factor $\sim13$ smaller
statistics in the data, both the significance level and their variations are
smaller than those in the Monte Carlo sample by a factor close to $\sqrt{13}$.

\begin{figure}
\begin{center}
\includegraphics[width=6.3in]{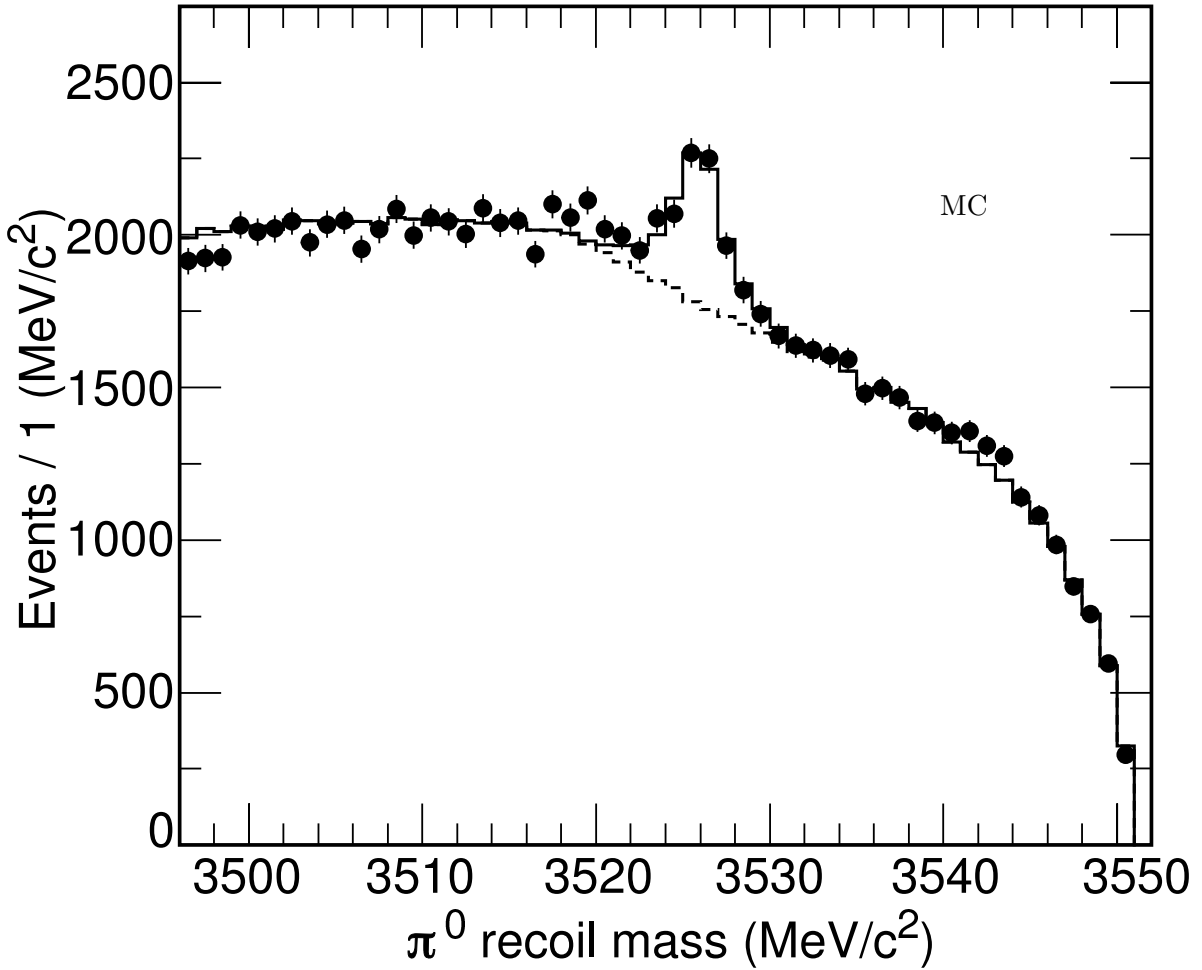}
\end{center}

\vspace*{-11.3cm}

~~~~~~~~~~~~~~~~~~~~~~~~~~~~~~~~~~~~~~~~~~~~~~~~~~~~~~~~~~~~~~~~~~~~~~~~~~~~~~MC

\vspace*{10.5cm}

\caption{Spectrum of masses (in GeV/$c^2$) recoiling against $\pz$ in a sample
of 39 million generic Monte Carlo events plus 15600 signal Monte Carlo events
($\ege1$-range inclusive analysis).  The solid histogram illustrates the fit
described in the text.
\label{fig:mchuge}}
\end{figure}

The $\pi^0$ recoil mass distribution for the Monte Carlo sample in the
$\ege1$-range analysis is shown in Figure \ref{fig:mchuge}. It was fitted using
the sum of two Gaussians with widths fixed to values determined by the
signal Monte Carlo sample.  The background was fitted using a histogram of the
$\pz$ recoil distribution from the generic Monte Carlo as described above.  The
dashed line shows the contribution of background without signal.

In the $M(\ec)$-range analysis, widths in $M(\ec)$ were determined by fits
using a Gaussian plus a low-order polynomial, while fits to $M(\hc)$ used a
Breit-Wigner resonance function with $\Gamma = 1$ MeV convolved with two
Gaussians, a quadratic polynomial
constrained to vanish at the kinematic endpoint, and an ARGUS background
function.

The best range of $\ec$ masses for optimizing signal significance was
determined via Monte Carlo studies using a likelihood ratio criterion.  Five
$M(\ec)$ windows 2940--3020, 2945--3015, 2950--3010, 2955--3005, and
2960--3000 MeV/$c^2$ were considered.  Upper and lower bounds were chosen
symmetrically with respect to $M(\ec) \simeq 2980$ MeV/$c^2$.  Detection of the
correct candidate for the E1 photon but assignment of a background $\pz$ with
the wrong energy as a signal $\pz$ candidate can introduce a potential
bias on $M(\hc)$ in the presence of asymmetric $M(\ec)$ limits.

Selecting events within the above $M(\ec)$ windows, fits were performed for
3496 MeV/$c^2 \le M(\hc) \le 3551.2$ MeV/$c^2$ to the generated
$h_c$ mass distributions.  The signal Monte Carlo was generated using a flat
angular distribution for the E1 photon.  A correction to the efficiency was
performed for the expected form $W(\cos \theta) \sim 1 + \cos^2 \theta$ with
respect to the beam axis. The ratio of the two efficiencies when integrating to
a maximum $|\cos \theta_{\rm max}|$ is $R_{\rm eff} = (1/4)(3 + \cos^2
\theta_{\rm max})$.  For $|\cos \theta_{\rm max}| = 0.804$, corresponding to
the outermost ring of the barrel calorimeter used in this analysis, the
correction factor is $R_{\rm eff} = 0.912$. The efficiencies were corrected for
$R_{\rm eff}$.

After fits to the signal Monte Carlo yielded the parameters of its Breit-Wigner
plus Gaussian functions, the generic Monte Carlo distribution was combined with
a weighted signal distribution to emulate a combined branching ratio for the
decay $\pp \to \pz \hc$ followed by $\hc \to \gme1 \ec$ of $\brp \brh = 4
\times 10^{-4}$.  The resulting distribution was fitted both with (generic +
weighted signal), and with generic background alone, yielding a ratio of
likelihoods.

\begin{figure}
\includegraphics[width=0.95\textwidth]{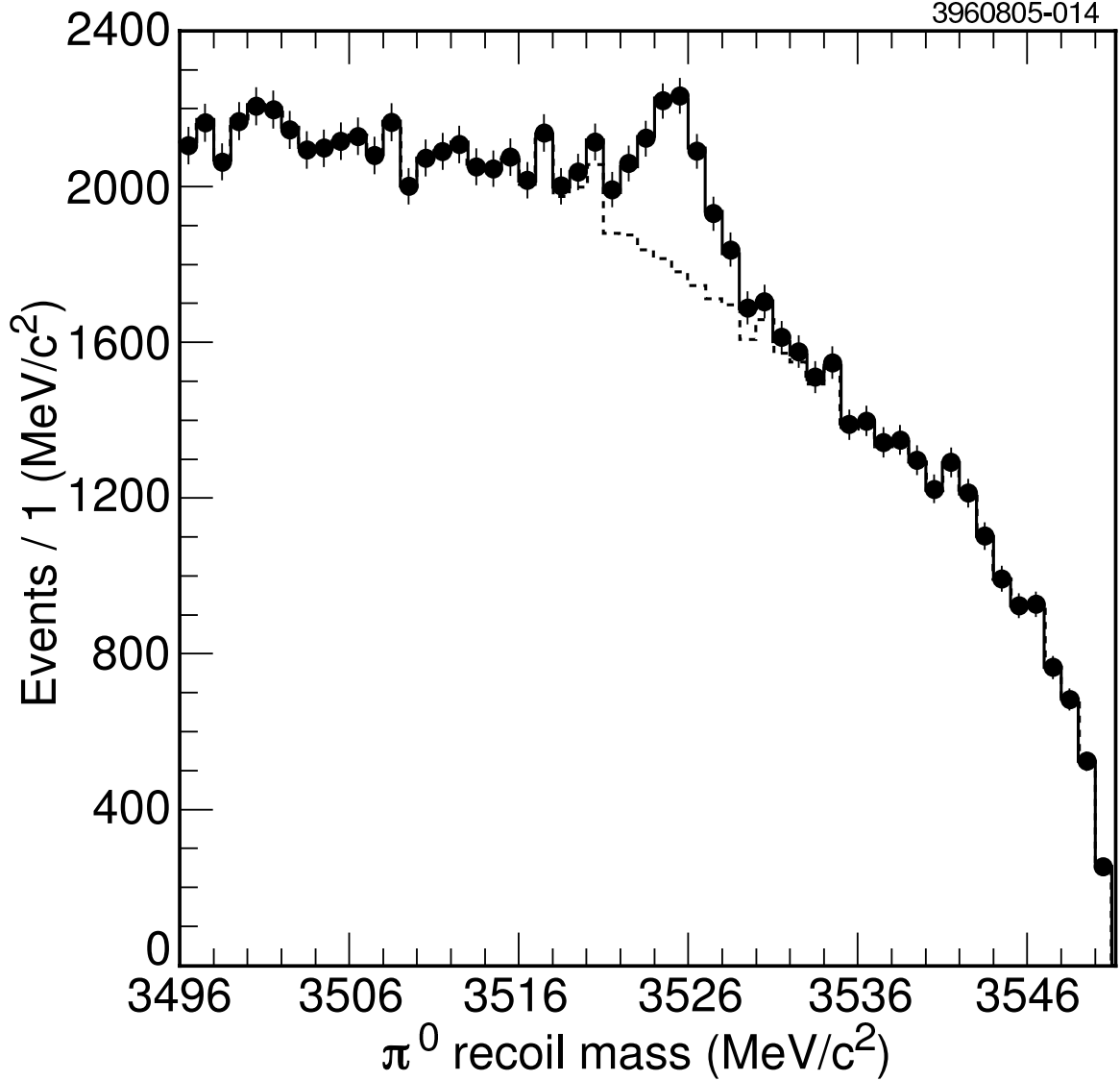}
\caption{Generic Monte Carlo $M(h_c)$ distribution ($M(\ec)$-range inclusive
analysis) for simulated $\pp$ data
of 39 million events with a signal of $69.7 \times 10^3$ $\hc$ decays
corresponding to $15.8 \times 10^3$ events of $h_c \to \gamma \eta_c$ for
2945 MeV/$c^2 \le M(\eta_c) \le 3015$ MeV/$c^2$.  The generated masses were
$[M(\eta_c),M(h_c)] = (2981.8,3525.3)$ MeV/$c^2$.  The signal was emulated
using a pair of Gaussians and a Breit-Wigner with $\Gamma = 1$ MeV.  The
dashed line shows the contribution of background.
\label{fig:mc}}
\end{figure}

This process resulted in an optimum range of 2945~MeV/$c^2 \le M(\ec) \le 3015$
MeV/$c^2$.  The corresponding $M(\hc)$ distribution is shown in Fig.\
\ref{fig:mc}.  For any wider $M(\ec)$ range, the photons from the transition
$\hc \to \gamma \ec$ become contaminated with contributions of
Doppler-broadened photons from the E1 transition $\chi_{c2}(3556) \to \gamma
\jp$.  Backgrounds from this transition and others rise steeply as the
upper limit on $M(\ec)$ is increased above 3020 MeV/$c^2$.

\begin{table}
\caption{Fits to simulated signal and background using a Breit-Wigner signal
function convolved with a double Gaussian and a generic Monte Carlo background
($M(\ec)$-range analysis).  Branching ratios include an efficency factor
$R_{\rm eff} =0.912$ for the $1 + \cos^2 \theta$ distribution of the E1 photon.
The nominal $M(\ec)$ range is labeled by an asterisk (*).
\label{tab:MCcomp}}
\begin{center}
\begin{tabular}{|c|c|c|c|c|c|} \hline \hline
 & \multicolumn{5}{c|}{$M(\eta_c)$ range (MeV/$c^2$)} \\ \cline{2-6}
 & 2940--3020 & *2945--3015 & 2950--3010 & 2955--3005 & 2960--3000 \\ \hline
 $M(h_c)({\rm MeV}/c^2)$ & 3525.24$\pm$0.16 & 3525.23$\pm$0.16
 & 3525.22$\pm$0.17 & 3525.21$\pm$0.17 & 3525.18$\pm$0.18 \\
Significance $\sigma$ & 17.08 & 17.30 & 17.20 & 17.05 & 16.45 \\
Efficiency (\%) & 15.3 & 14.6 & 13.5 & 12.2 & 10.6 \\
$\brp \brh \times10^{-4}$ & $4.07\pm0.25$ & $4.07\pm0.25$ & $4.07\pm0.25$
 & $4.07\pm0.25$ & $4.07\pm0.26$ \\ \hline \hline
\end{tabular}
\end{center}
\end{table}

Fits to simulated signal and background in the $M(\ec)$-range analysis are
compared in Table \ref{tab:MCcomp}.  The $\eta_c$ mass range 2945--3015
MeV/$c^2$ gives the greatest signal significance for an $h_c$ of mass 3525.3
MeV/$c^2$ produced with ${\cal B}(\pp \to \pi^0 h_c){\cal B}(h_c \to \gamma
\eta_c)=4 \times 10^{-4}$.  The extracted
values of $M(h_c)$ are about 0.1 MeV/$c^2$ below the input. This feature is
included in the estimate of systematic errors.  The maximum significance of
$17.3 \sigma$ scales to $4.8 \sigma$ for a sample of $3.08 \times 10^6$ events.
\bigskip

\leftline{\it 3.  Variations in output parameters.}

In the generic Monte Carlo sample, for all the 22 variations of the
$\ege1$-range analysis listed in Table
\ref{tab:summ}, the change in output $M(h_c)$ and $\brp \brh$ were found to
be $\Delta M(h_c) \le 0.1$ MeV/$c^2$, and $\Delta(\brp \brh) \le 0.2
\times 10^{-4}$, i.e., within the statistical errors assigned by the output.
To see the level of statistical variations in Monte Carlo
samples as small as the data (i.e., $\sim3$ million $\pp$),
the total sample of 39~million $\pp$~decays  was split into 13~independent
samples, each of 3 million $\pp$.  Table \ref{tab:smalla}
summarizes results of the analysis for the choices of the final
selection and for variations of these choices.  For the final selection
the limits of variation were found to be $\Delta M = (-0.4,+0.3)$ MeV/$c^2$ and
$\Delta (\brp \brh) = (-1.1,+1.4) \times 10^{-4}$.  For $\brp
\brh$ the effect of variations from the final selection is within the
range observed for the final selection.  There may be some
evidence of larger than expected variation when one changes
$\Delta \ege1$ to $\pm50$ MeV, and when one includes more than one signal
$\pz$ candidate.  A choice of $\Delta \ege1 = 50$ MeV begins to accept photons
on the high-energy tail of the transition $\chi_{c2} \to \gamma J/\psi$ when
detector resolution and recoil effects are taken into account.

Because the Monte Carlo signal sample was generated with an assumed $M(\hc) =
3526$ MeV/$c^2$, or $\ege1 =503$ MeV, it is prudent to examine what bias is
introduced in $M(\hc)$ and $\brp \brh$ if the true $M(\hc)$ were to differ
from 3526 MeV/$c^2$.  The resulting variation in efficiency
was found to be less than 2.5\% for $M(h_c) = 3526 \pm 14$ MeV/$c^2$.

\begin{table}
\caption{Results for $M(h_c)$ and $\brp \brh$ from trial experiments with 13
independent Monte Carlo samples of 3 million $\pp$ each [$\ege1$-range
analysis].  The inputs were $M(h_c)=3526.0$ MeV/$c^2$ and $\brp \brh = 4.0
\times 10^{-4}$.  The full Monte Carlo sample yielded $M(h_c) = 3526.1 \pm 0.1$
MeV/$c^2$ and $\brp \brh = 4.1 \pm 0.3 \times 10^{-4}$.
Variations from the final selection resulted in $\Delta M(h_c) \le
0.1$ MeV/$c^2$ and $\Delta (\brp \brh) \le 0.2 \times 10^{-4}$ for this large
sample.  The second column lists $\Delta M(h_c)\equiv M(h_c)-3526$ MeV/$c^2$
or $\Delta (\brp \brh)\equiv (\brp \brh) - 4.0 \times 10^{-4}$ for the
final selection.  The following columns list $\Delta M(h_c)$ or $\Delta (\brp
\brh)$ for the specified variations from the final selection.  The statistical
error on all output masses was $\pm0.5$ to $\pm0.6$ MeV/$c^2$ and on all output
$\brp \brh$ was $1.0 \times 10^{-4}$.
\label{tab:smalla}}
\begin{center}
\begin{tabular}{|c|c|c|c|c|c|} \hline \hline
 & $\Delta M(h_c)$ -- MeV/$c^2$ &  \multicolumn{4}{c|}{$\Delta M(h_c)$
-- MeV/$c^2$ with variations from final} \\ \cline{3-6}
 & Final selection &  $\Delta E_\gamma \pm50$ MeV & $\ge 1 \pz$ & No endcap
 & No $\eta$ supp.\ \\ \hline
MC & --0.4/+0.3  & --1.8/+0.7 & --2.1/+0.2 & --0.3/+0.2 & --0.4/+0.3\\
Data &  &  +0.4  &      +0.4  &    +0.4    &    +0.2\\ \hline
 & $\Delta (\brp \brh \times 10^{4})$ &
 \multicolumn{4}{c|}{$\Delta (\brp \brh \times 10^{4})$ -- with variations
 from final} \\ \cline{3-6}
 & Final selection &  $\Delta E_\gamma \pm50$ MeV & $\ge 1 \pz$ & No endcap
 & No $\eta$ supp.\ \\ \hline
MC & --1.1/+1.4  & --1.2/+0.7 & --0.3/+1.1 & --0.5/+0.3 & --1.1/+0.3\\
Data & &    --0.5    &     --0.5  &    +0.0   &   --0.4 \\ \hline \hline
\end{tabular}
\end{center}
\end{table}

\begin{table}
\caption{Results for $M(h_c)$ and $\brp \brh$ from trial experiments with 13
independent Monte Carlo samples of 3 million $\pp$ each [$M(\ec)$-range
analysis].  The inputs were $M(h_c)=3525.3$ MeV/$c^2$ and $\brp \brh
= 4.0 \times 10^{-4}$.  The full Monte Carlo sample yielded $M(h_c) = 3525.33
\pm 0.18$ MeV/$c^2$ and $\brp \brh = 3.9 \pm 0.3 \times 10^{-4}$.  The second
column lists $\Delta M(h_c)\equiv M(h_c)-3525.3$ MeV/$c^2$ or $\Delta (\brp
\brh)\equiv (\brp \brh) - 4.0 \times 10^{-4}$ for the final selection.  The
following columns list $\Delta M(h_c)$ or $\Delta (\brp \brh)$ for variations
from the final selection. 
\label{tab:smallb}}
\begin{center}
\begin{tabular}{|c|c|c|c|c|c|c|} \hline \hline
 & $\Delta M(h_c)$ (MeV/$c^2$) & \multicolumn{5}{c|}{$\Delta M(h_c)$
(MeV/$c^2$) with variations from final} \\ \cline{3-7}
 & Final selection & $\Delta M(\ec) \pm40$ MeV & $\Delta M(\ec) \pm20$ MeV
 & $\ge 1 \pz$ & w/endcap & w/$\eta$ supp.\ \\ \hline
MC & --0.5/+0.3 & --0.5/+0.3 & --0.5/+0.4 & --0.4/+0.3 & --0.4/+0.4
 & --0.4/+0.5 \\
Data & & +0.4 & --0.3 & +0.0 & +0.5 & --0.4 \\ \hline
 & $\Delta (\brp \brh \times 10^{4})$ &
 \multicolumn{5}{c|}{$\Delta (\brp \brh \times 10^{4})$ -- with variations
 from final} \\ \cline{3-7}
 & Final selection & $\Delta M(\ec) \pm40$ MeV & $\Delta M(\ec) \pm20$ MeV
 & $\ge 1 \pz$ & w/endcap & w/$\eta$ supp.\ \\ \hline
MC & --0.7/+0.5 & --0.7/+0.4 & --0.6/+0.4 & --0.6/+0.3 & --0.5/+0.4
 & --0.5/+0.6 \\
Data & &  --0.6   & +0.1  &  --0.3  & --1.5 & +0.1 \\ \hline \hline
\end{tabular}
\end{center}
\end{table}

The corresponding variations in the $M(\ec)$-range analysis were explored by
again forming 13 samples of $\sim 3$ million generic $\pp$ Monte Carlo and
adding 13
samples of 3135 signal Monte Carlo events with ${\cal B}(h_c \to\gamma
\eta_c) = 37.7\%$, ${\cal B}(h_c \to ggg) = 56.8\%$, and ${\cal B}(h_c \to
\gamma g g) = 5.5\%$.  This permitted simulation of a combined branching ratio
$\brp \brh = 4 \times 10^{-4}$.  Fits were performed using the same functions
used in fitting data.  The results are shown in Table \ref{tab:smallb}.
Deviations from the mean were found to be of the expected magnitude for data
samples of this size.
\bigskip

\leftline{\it 4.  Quality of generic Monte Carlo simulation.}

Because the CLEO generic Monte Carlo is used to determine optimum selection
criteria
for energy ranges and binary choices, one must quantify its level of agreement
with data in emulating the $M(\hc)$ spectrum.  The EvtGen \cite{EvtGen}
generator is combined with a JETSET \cite{JETSET} version tuned to match the
relevant low-energy regime \cite{Athar:2004dn}.  For photon energies below
450~MeV and pion momenta below 550~MeV/$c$, the data and Monte Carlo agree
within $\pm5\%$.  Above these values the ratio of data to Monte Carlo falls
below 95\%, rising again from $\sim
90\%$ above $E_\gamma = 600$ MeV and from $\sim 85\%$ above $p(\pz) = 950$
MeV/$c$.  For low energy photons in the slow $\pi^0$ from $\pp \to \pz \hc$,
the generic Monte Carlo is satisfactory, but its use over extended ranges of
energy and momenta, as required in determining background shapes, may not be
so.  This provides a motivation for basing the background shapes on the
data, i.e., the $\pi^0$ recoil spectrum without requiring $E_\gamma = 503 \pm
35$~MeV, instead of the $\pi^0$ recoil spectrum from the generic Monte Carlo.
\bigskip

\leftline{\it 5. Choices in $M(\ec)$ analysis.}

\begin{table}[t]
\caption{Binary choices of selection and criteria ($M(\ec)$ analysis).
Asterisks denote nominal choices.
\label{tab:bincomp}}
\begin{center}
\begin{tabular}{|c|c|c|c|c|c|c|} \hline \hline
 $\ggJ$/ & Range & $\eta$ & MC & \multicolumn{3}{c|}{Signal} \\
\cline{5-7}
  mTk & & supp & $\sigma$ & Mass (MeV/$c^2$) & Evts.\ in pk.\
 & $\br~(10^{-4})$ \\ \hline                                   
 *$\ggJ$ & *$M(\ec)$ & *No & 17.3 & $3525.3\pm0.6$ & $159\pm41$ 
 &$3.5\pm0.9$ \\ \hline 
 *$\ggJ$ & *$M(\ec)$ & Yes &  16.9 &  $3524.9\pm0.6$ & $132\pm35$ 
 & $3.6\pm1.0$ \\ \hline
 *$\ggJ$ & $\ege1$ & *No &  16.7 & $3525.3\pm0.7$ & $161\pm44$ 
 & $3.4\pm0.9$ \\ \hline
 *$\ggJ$ & $\ege1$ & Yes & 16.3 & $3524.8\pm0.6$ & $134\pm37$ 
 & $3.6\pm1.0$ \\ \hline
    mTk   & *$M(\ec)$ & *No & 17.3 & $3525.1\pm0.6$ & $152\pm42$ 
 & $3.3\pm0.9$ \\ \hline
    mTk   & *$M(\ec)$ & Yes  & 16.9 & $3524.7\pm0.6$ & $134\pm36$ 
 & $3.6\pm1.0$ \\ \hline
    mTk   & $\ege1$ & *No & 16.6 & $3525.1\pm0.7$ & $145\pm41$ 
 & $3.1\pm0.9$ \\ \hline
    mTk   & $\ege1$ & Yes & 16.2 & $3524.7\pm0.5$ & $136\pm38$ 
 & $3.6\pm1.0$ \\ 
 \hline \hline
\end{tabular}
\end{center}
\end{table}

In the $\ege1$ analysis, electromagnetic cascades involving E1
transitions to and from intermediate $\chi_c$ states were suppressed by
excluding events with the effective mass of charged tracks exceeding 3050
MeV/$c^2$ (``mTk'' criterion). In the $M(\ec)$ analysis, the mass
recoiling against $\gamma \gamma$ was reconstructed directly (``$\gamma
\gamma$'' criterion), and events with a recoil mass within $\pm 40$ MeV/$c^2$
of $M(J/\psi)$ were excluded.

In the $M(\ec)$ analysis, which does not use endcap photons and does
not restrict photons in $\pz$ candidates to the ten most energetic showers,
an advantage in Monte Carlo significance by about $0.6\sigma$ appears when
the $M(\ec)$ range rather than the $\ege1$ range is selected.

In the $\ege1$ analysis, Monte Carlo likelihood ratios favor suppressing
$\gme1$ candidates which can form an $\eta$ when paired with other photons.
In the $M(\ec)$ analysis, which uses a larger pool of photon candidates for
possible pairings, such a suppression entails a loss of efficiency for
signal detection, leading to decreased significance in Monte Carlo by $0.4
\sigma$.  The $M(\ec)$ analysis consequently does not adopt this suppression.

The above three criteria were compared in a binary manner, leading to the
results shown in Table \ref{tab:bincomp}.  The effects of each variation are
largely independent of each other when measured by change in significance.
The first row was chosen over the fifth in the $M(\ec)$ analysis on the basis
of a very slight excess in Monte Carlo (MC) significance $\sigma$;
differences in resulting mass and branching ratio are within statistics.

\bigskip
\leftline{\it 6.  Dependence on branching ratio $\br(\hc \to \gme1 \ec)$ and
$M(\hc)$ in signal Monte Carlo.}

In the $\ege1$ analysis, Monte Carlo simulations were performed by assuming
$\brh \equiv \br(\hc \to \gme1 \ec) = 100\%$ rather than the value of 37.7\% 
\cite{Godfrey:2002rp} used in the $M(\ec)$ analysis. Moreover, slightly
different values of $M(\hc)$ for the signal Monte Carlo were used in the two
analyses.  The results of changing just $\brh$ or both $\brh$ and $M(\hc)$ in
the signal Monte Carlo were studied for the $M(\ec)$ analysis.
Several features were notable in this comparison.

(1) The maximum signal likelihoods in Monte Carlo were less for the choice of
$\brh = 100\%$:  (15.5,16.1)$\sigma$ for $M(\hc) = (3525.3,3526.0)$ MeV/$c^2$
versus 17.3$\sigma$ for $\brh = 37.7\%$ and $M(\hc) = 3525.3$ MeV/$c^2$.
(2) For the same $M(\ec)$ range, the values of $M(\hc)$ in data were stable
under variation of $\brh$ or input $M(\hc)$, while the extracted values of
$\brp \brh$ rose by about $0.4 \times 10^{-4}$ when $\brh = 100\%$ was taken
in the signal Monte Carlo.
(3) When $\brh = 100\%$, the maximum signal likelihood in Monte Carlo still
favored no $\eta$ suppression applied to the E1 photon, but to a lesser extent.

Because the variations in $M(\hc)$ and $\brp \brh$ observed under the above
changes were ascribable to the signal fitting hypothesis rather than to the
data themselves, they were included in estimates of systematic error,
giving $\Delta M(\hc) = -0.1$ MeV/$c^2$ and $\Delta \brp \brh = + 0.4
\times 10^{-4}$.
\bigskip

\leftline{\it 7.  Asymmetric $M(\ec)$ selection windows.}

The $\ec$ mass windows were chosen symmetric about 2980 MeV/$c^2$ in the
$M(\ec)$ analysis to avoid $M(\hc)$ spectrum distortions
if an E1 photon of the correct energy were paired with a random
pion not associated with the transition $\pp \to \pz \hc$.  Slightly higher
Monte Carlo significance (17.5$\sigma$ versus nominal 17.3$\sigma$) occurs with
the asymmetric window 2955--3015 MeV/$c^2$ (versus nominal 2945--3015
MeV/$c^2$).  On the other hand, the signal significance in data peaks for the
asymmetric window 2945--3005 MeV/$c^2$ at $4.6\sigma$ (versus
4.0$\sigma$ for the nominal window), and the value of $M(\hc)$ obtained from
the data is 0.4 MeV/$c^2$ lower.  This behavior is
consistent with the lower $\ec$ masses observed in a recent analysis of
$\pp$ radiative decays \cite{Athar:2004dn} and in the exclusive analysis
reported below.
\bigskip

\leftline{\bf B.  Exclusive analysis}
\bigskip

The signal Monte Carlo indicates that the reconstructed (or recoil) $\ec$ mass
and width are mode dependent because of the different final decay particles.
The value of $M(\ec)$
calculated after kinematic fitting was required to be within 50 MeV/$c^2$ of
the nominal mass.  Monte Carlo events indicate that this is more than 80\%
efficient. The width of the reconstructed $\ec$ mass distribution depends on
both the detector resolution and the intrinsic width, $\gec$. The latter
has not been well measured \cite{Eidelman:2004wy,etac}, and the former is
decay-mode dependent.  Because the
requirement that $M(\ec)$ be within 50 MeV/$c^2$ of its nominal value is loose,
the systematic uncertainty of the efficiency due to this requirement is minimal,
however.  Measuring the ratio of branching ratios for cascade decay and direct
radiative decay reduces this systematic uncertainty further.  In addition to
the other criteria in Table III, this analysis takes the $\pz$ pull mass limit
for signal selection and $\pz$ suppression to be 3, and the reduced $\chi^2$ for
kinematic constraints to be less than 10. The direct radiative decay $\direct$
is studied in the same $\ec$ decay modes, using similar event selection criteria
except that the $M(\ec)$ and signal $\pi^0$ selection criteria are dropped, and
the $\ec$ yield is determined from the fit to the $\gamma$ recoil mass
spectrum.

\newpage
\centerline{\bf VII.  THE SIGNAL IN THE DATA}
\bigskip

\leftline{\bf A.  Inclusive analyses}
\bigskip

Figure \ref{fig:orig} shows the spectrum of recoils against $\pi^0$ for the
data in Table \ref{tab:data} with the event selection criteria determined to
optimize the signal sensitivity in the $\ege1$ analysis.  These data were
fitted with background as determined in Sec.\ VI plus a Breit-Wigner resonance
of width 0.9 MeV.  The background used was the $\pi^0$ recoil spectrum without
the cut on $\ege1$. The peak shape
consisted of the Breit-Wigner width convolved with an instrumental resolution
function, determined from the signal Monte Carlo simulation, which itself was
fitted with a double Gaussian.  The efficiency for the final event selection
was determined to be $\epsilon=13.4\%$. The results are:

\begin{figure}
\begin{center}
\includegraphics[width=6.3in]{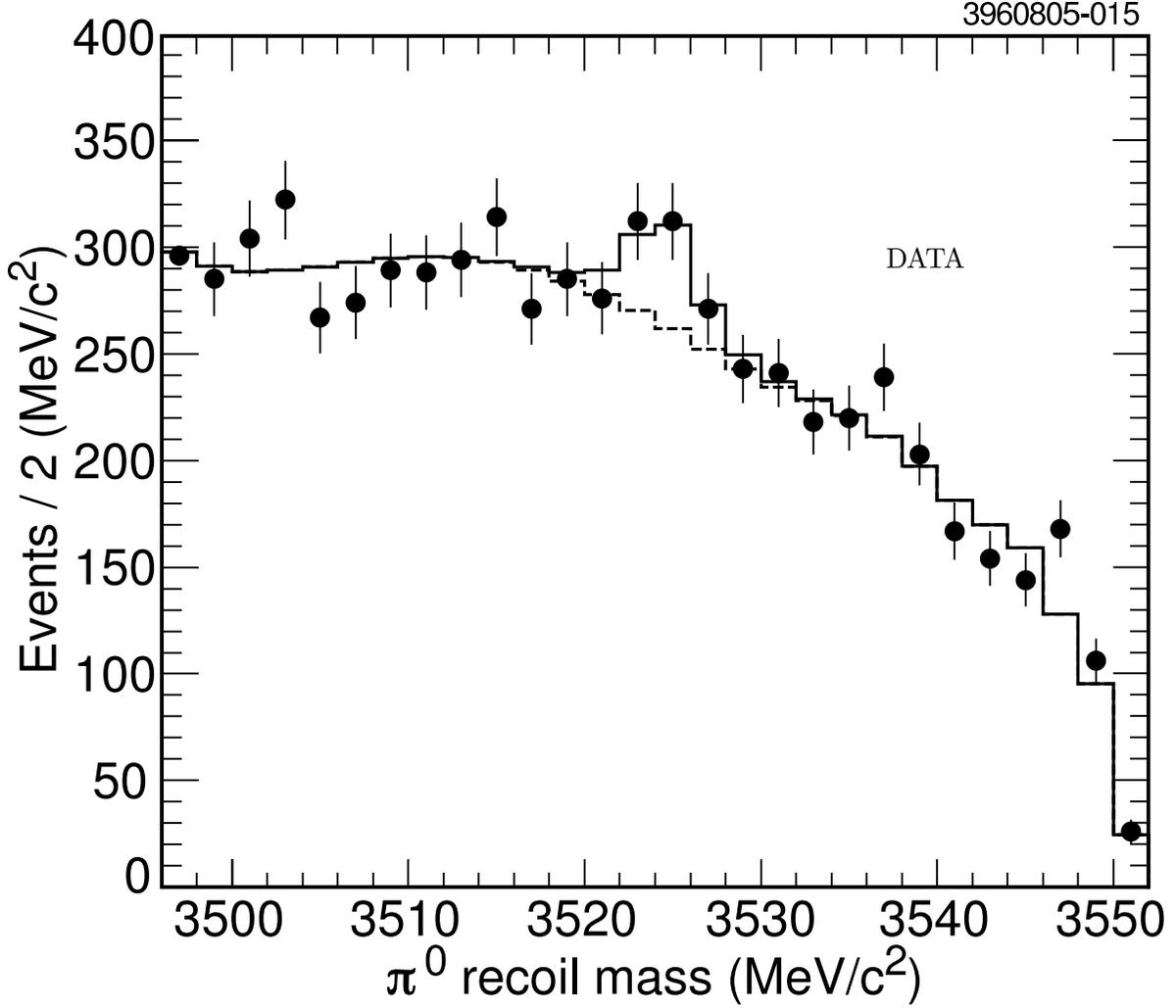}
\end{center}
\vspace*{-11.3cm}

~~~~~~~~~~~~~~~~~~~~~~~~~~~~~~~~~~~~~~~~~~~~~~~~~~~~~~~~~~~~~~~~~~~~~~~~~~~DATA
\vspace*{10.0cm}

\caption{$M(h_c)$ distribution from recoil $\pi^0$ for the CLEO III + CLEO-c
data
set corresponding to the final event selection in inclusive analysis based on
selecting a range of $\ege1$.  The dashed line denotes the background function.
The $\chi^2$ per degree of freedom for the fit including peak and background is
34.1/25 = 1.36, as noted in Table \ref{tab:summ}.  The corresponding confidence
level is 10.5\%.
\label{fig:orig}}
\end{figure}

\begin{itemize}

\item
N(evts) = $139\pm41$, \quad significance = $3.6\sigma$

\item
$M(h_c)=3524.4\pm 0.7$ MeV/$c^2$

\item
$\brp \brh \equiv {\cal B}(\pp \to \pi^{0} h_c) \times
{\cal B}(h_c \to \gamma\eta_c)$ = (3.4$\pm$1.0)$\times 10^{-4}$.

\end{itemize}

When selecting a range of $M(\ec)$from Monte Carlo, choosing events in the
interval 2945--3015 MeV/$c^2$ gave the greatest signal significance, and hence
this interval was used for further analysis.
For the data the significance is slightly greater for a
narrower range of $M(\ec)$, as shown in Table \ref{tab:DATAcomp}.
The resulting $h_c$ mass spectrum is shown in Fig.\ \ref{fig:nomdt}.
The results are: 
\begin{figure}
\includegraphics[height=5.5in]{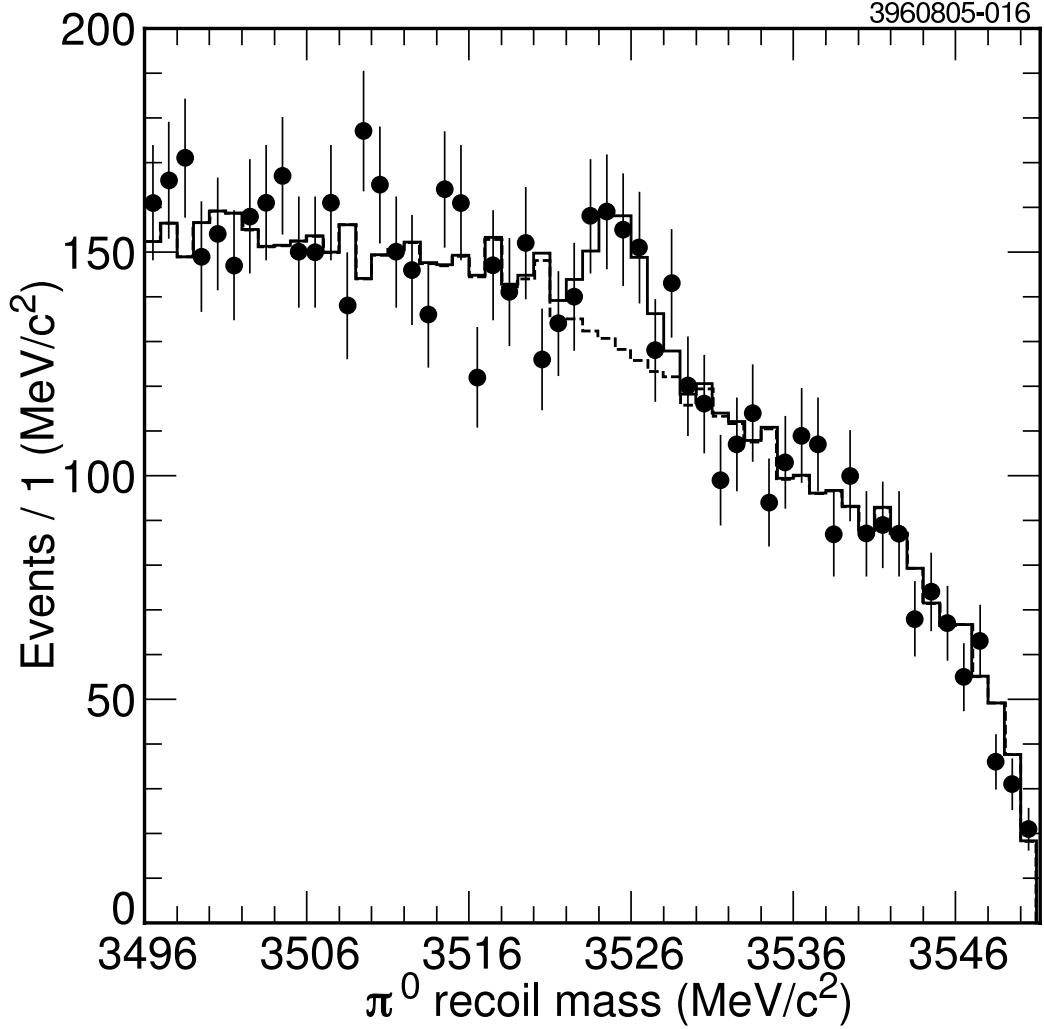}
\caption{$M(h_c)$ distribution from recoil $\pi^0$
for 2945 MeV/$c^2 \le M(\eta_c) \le 3015$
MeV/$c^2$, fitted over the range 3496 MeV/$c^2 \le M(h_c) \le 3551.2$ MeV/$c^2$
[analysis selecting range of $M(\ec)$].  The curve denotes the
background function based on generic Monte Carlo plus a signal as described
in Sec.\ VI B.  The dashed line shows the contribution of background
alone.  The peak contains $159 \pm 41$ events.  The confidence level of the
fit to signal + background was 34\%, corresponding to $\chi^2 = 55.6$ for
52 degrees of freedom. 
\label{fig:nomdt}}
\end{figure}
\begin{itemize}

\item
N(evts) = $159\pm41$, \quad significance = $4.0\sigma$

\item
$M(h_c)=3525.3\pm 0.6$ MeV/$c^2$

\item
$\brp \brh = (3.5 \pm 0.9) \times 10^{-4}$.

\end{itemize}
The CLEO-III and CLEO-c data were fitted separately. Results are shown in Table
\ref{tab:summ} for the $\ege1$ analysis and Table \ref{tab:IIIvsC} for the
$M(\ec)$ analysis.  The relative weights of the two samples [with values of
$M(\hc)$ differing by about 2 MeV/$c^2$] differ between the two analyses, with
the $\ege1$ analysis finding fewer signal events in the CLEO-c sample while the
$M(\ec)$ analysis finds approximately equal signals in the CLEO III and CLEO-c
samples.  This accounts for the major part of the difference between $M(\hc)$
values in the combined samples.
No such difference was found in Monte Carlo simulations of CLEO-c data,
indicating that the observed difference is purely statistical.
\bigskip

\begin{table}
\caption{Same as Table \ref{tab:MCcomp} for fits to CLEO-III and CLEO-c $\pp$
data [$M(\ec)$ analysis].
\label{tab:DATAcomp}}
\begin{center}
\begin{tabular}{|c|c|c|c|c|c|} \hline \hline
 & \multicolumn{5}{c|}{$M(\eta_c)$ range (MeV/$c^2$)} \\ \cline{2-6}
 & 2940--3020 & *2945--3015 & 2950--3010 & 2955--3005 & 2960--3000 \\ \hline
 $M(h_c)({\rm MeV}/c^2)$ & 3525.67$\pm$0.85 & 3525.26$\pm$0.60
 & 3525.08$\pm$0.55 & 3525.06$\pm$0.57 & 3524.97$\pm$0.58 \\
Signif.\ $\sigma$ & 3.24 & 4.03 & 4.27 & 4.22 & 3.97  \\
$\brp \brh \times 10^4$ & $2.86\pm0.91$ & $3.53\pm0.91$ & $3.76\pm0.92$ &
 $3.76\pm0.93$ & $3.65\pm0.97$ \\ \hline \hline
\end{tabular}
\end{center}
\end{table}

\begin{table}
\caption{$M(\hc)$ and combined branching ratio $\brp \brh$ for separate
CLEO-III and CLEO-c data samples [$M(\ec)$ analysis, range 2945--3015
MeV/$c^2$].
\label{tab:IIIvsC}}
\begin{center}

\begin{tabular}{|c|c|c|c|} \hline \hline
Data     &     Mass        &  Events   &    Branching      \\
sample   &  (MeV/$c^2$)    &  in peak  & ratio ($10^{-4}$) \\ \hline
CLEO-III &  $3524.1\pm1.0$ & $86\pm29$ & $3.8\pm1.3$ \\
CLEO-c   &  $3526.6\pm0.8$ & $93\pm29$ & $4.2\pm1.3$ \\
\hline \hline
\end{tabular}

\end{center}
\end{table}

\begin{figure}
\includegraphics[height=4.3in]{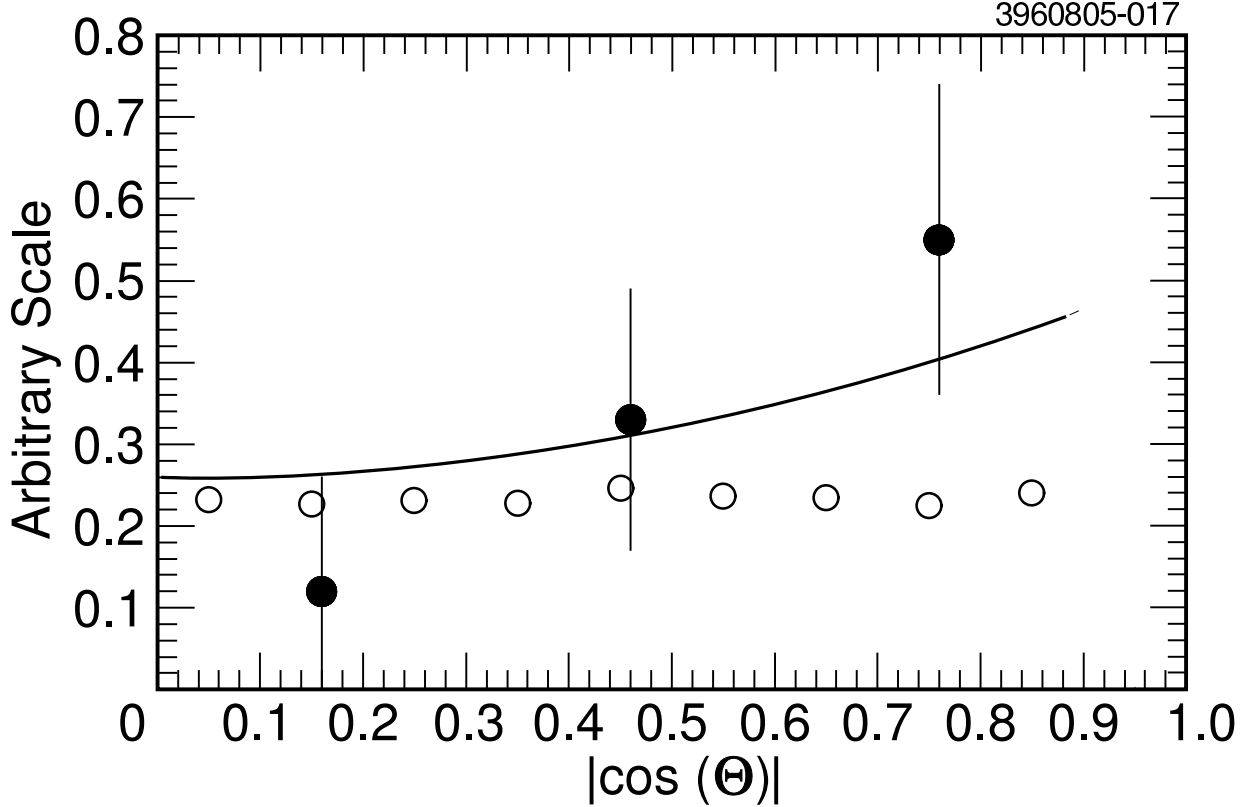}
\caption{Angular distribution of the photons with $E_\gamma = 503 \pm 35$ MeV
from the inclusive analysis.  Solid points denote yield of photons from
$\hc \to \gamma \eta_c$, while open circles denote background photons.  The
curve shows the fit of the $\hc \to \gamma \ec$ points with a $1 + \cos^2
\theta$ distribution.  The background photons are seen to be isotropically
distributed.  Scales for the three plots are arbitrary.
\label{fig:cos}}
\end{figure}

The angular distribution of the $\gme1$ photon in the inclusive analysis was
obtained by fitting separately the $\hc$ peak in the angular ranges $0.0
\le |\cos \theta| \le 0.3$, $0.3 \le |\cos \theta| \le 0.6$, and $0.6 \le
|\cos \theta| \le 0.9$.  The results are presented in Fig.\ \ref{fig:cos}.  A
$1 + \cos^2 \theta$ distribution, as expected for an E1 transition from a spin
1 state, gives a satisfactory fit, with $\chi^2 =1.7$ for 2 degrees of freedom.
The angular distribution for the background, obtained in the same way as for
the fit to the signal, corresponds to the dotted histogram in
Fig.\ \ref{fig:orig}, and is flat as expected.

\bigskip

\leftline{\bf B.  Exclusive analysis}
\bigskip

There are several ways to search for an $h_{c}$ signal in exclusive modes.
One may observe enhancements in the photon energy spectrum from $\hc \to \gamma
\ec$, the reconstructed $\hc$ mass spectrum, or the recoil $\pz$ energy
spectrum. The photon energy resolution $\sigma(E)/E$ is 2.1\% to 3.8\% for a
photon of energy around 500 MeV, depending on whether it is in the barrel or
endcap CsI~calorimeter.  The signal photon energy also has a spread because of
the intrinsic width of $\ec$. The reconstructed $\hc$ mass calculated from the
4-momenta of the $\eta_{c}$ and the transition photon also has poor resolution,
and depends on $\eta_{c}$ decay modes. In the signal Monte Carlo, both the
photon energy resolution and reconstructed $h_{c}$ mass resolution are larger
than 15 MeV in all modes used. The recoil $\pz$ (from $\pp\to \pz \hc$) has
much better energy resolution because of the $\pz$ mass constraint fit employed
in the $\pi^{0}$ reconstruction algorithm, as mentioned previously.
The $M(h_{c})$ spectrum recoiling against a $\pz$ is also independent of
$\ec$ decay modes, so one can fit the $h_{c}$ signal with the same signal shape
when signals from different modes are added together.

\begin{figure}
\begin{center}
\epsfig{file=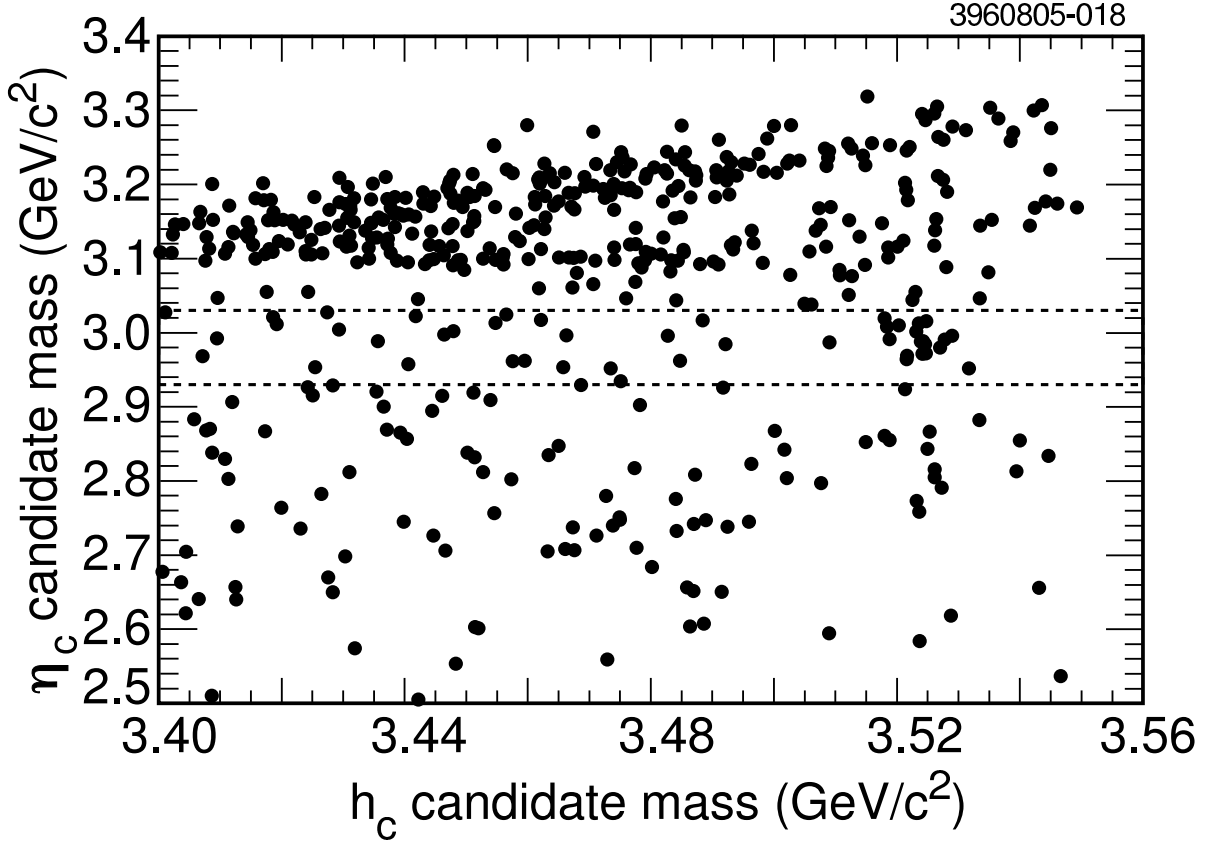,height=4.5in}
\caption{Scatter plot of the reconstructed $\eta_{c}$ mass versus the $\hc$
candidate mass obtained from $\pi^0$ recoil
in data for the exclusive analysis.  The
horizontal band near $M(J/\psi) = 3097$ MeV/$c^2$ and the diagonal band at
larger $\ec$
candidate mass correspond to $\pp \rightarrow \pi^0\pi^0 J/\psi$ and $\pp
\rightarrow \gamma \chi_{c0}$, respectively.  The dashed lines denote the
region $M(\eta_c) = 2982 \pm 50$ MeV/$c^2$.  In this band a cluster of events
is visible around $M(\hc) = 3524$ MeV/$c^2$.
\label{fig:hcetacmass}}
\end{center}
\end{figure}

After all the selection criteria except for $M(\ec)$ are imposed, there is a
clear cluster of events in the plot of $\ec$ candidate mass versus $\pz$ recoil
mass, shown in Fig.~\ref{fig:hcetacmass}.  Properties of the nineteen events
in the $M(\ec)$ band between the dotted lines and with $M(\hc)$ between
3516 and 3530 MeV/$c^2$ are summarized in Table \ref{tab:exclevts}.

\begin{table}
\caption{List of exclusive event candidates.
\label{tab:exclevts}}
\begin{center}
\begin{tabular}{c c c c c} \hline \hline
Mode    & $M(\hc)$    & $E^*_\gamma$ & \multicolumn{2}{c}{$M(\ec)$ (MeV/$c^2$)}
 \\
        & (MeV/$c^2$) & (MeV) & Reconstructed & Recoil \\ \hline
$\kskp$ & 3524.3 & 475.0 & 3018.7 & 3012.0 \\
        & 3529.3 & 496.4 & 2995.3 & 2991.9 \\ \hline
$\klkp$ & 3521.7 & 513.4 &   --   & 2964.2 \\
        & 3521.5 & 541.2 &   --   & 2930.8 \\
        & 3517.7 & 463.2 &   --   & 3019.2 \\
        & 3523.5 & 486.1 &   --   & 2998.3 \\ \hline
$\kkpp$ & 3525.0 & 499.9 & 2989.2 & 2983.4 \\
        & 3524.3 & 474.5 & 2978.8 & 3012.7 \\
        & 3526.7 & 507.1 & 2989.5 & 2976.8 \\ \hline
$\fourpi$ & 3527.2 & 494.1 & 2983.3 & 2992.6 \\
        & 3520.4 & 475.9 & 2975.3 & 3007.1 \\
        & 3523.0 & 471.6 & 2987.5 & 3014.8 \\
        & 3530.9 & 523.0 & 2956.5 & 2962.0 \\
        & 3519.2 & 498.7 & 2992.6 & 2979.0 \\
        & 3519.8 & 463.2 & 3009.1 & 3021.3 \\
        & 3524.0 & 473.8 & 3007.6 & 3013.2 \\
        & 3524.8 & 517.5 & 2972.5 & 2962.4 \\ \hline
$\kkpizero$ & 3525.4 & 497.7 & 2976.1 & 2986.5 \\ \hline
$\ppetatwo$ & 3521.1 & 414.4 & 3013.0 & 3078.8 \\ \hline \hline
\end{tabular}
\end{center}
\end{table}

\begin{figure}
\begin{center}
\epsfig{file=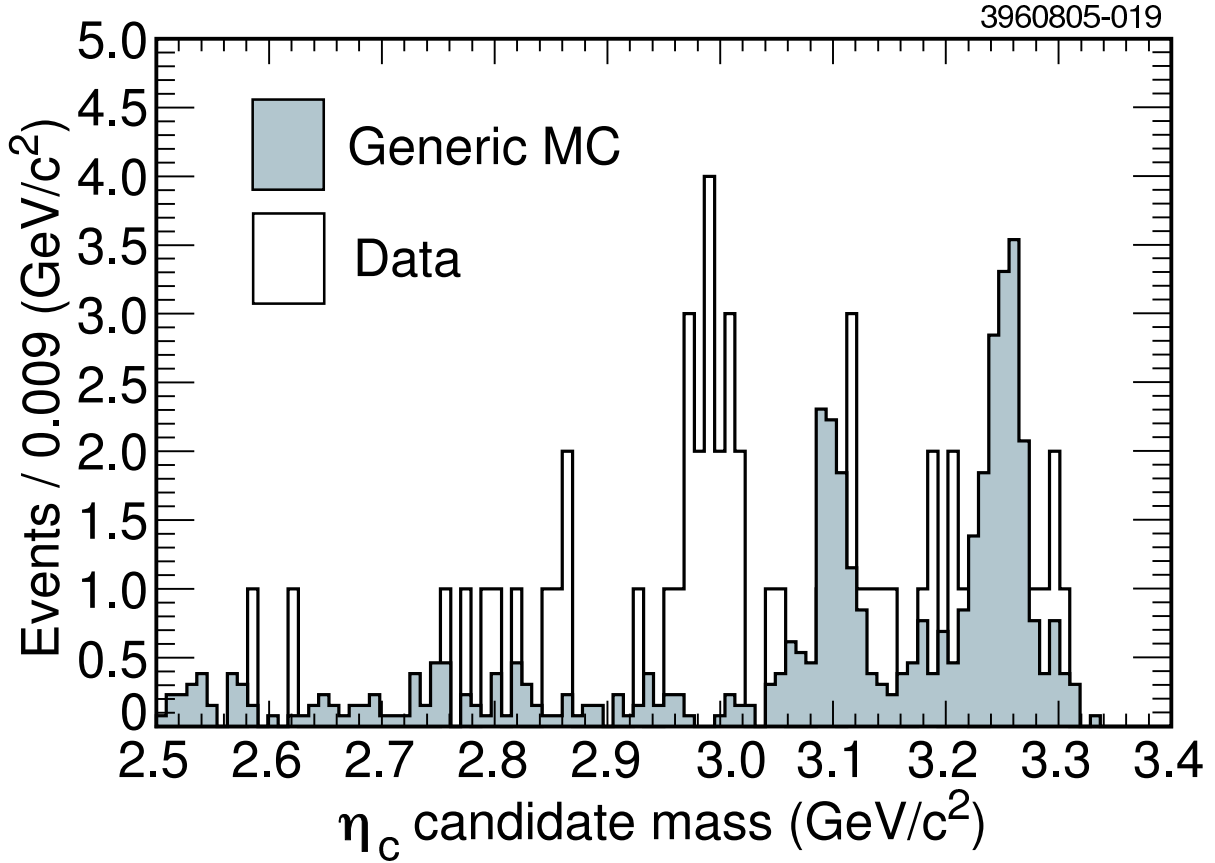,height=4.6in}
\caption{Data events (open histograms) and Monte Carlo background estimate
(shaded histograms) of reconstructed $\eta_c$ candidate mass projection for
$M(\pi^0\;{\rm recoil}) = 3524 \pm 8 $ MeV/$c^2$.
\label{fig:etacmass}}
\end{center}
\end{figure}

\begin{figure}
\begin{center}
\epsfig{file=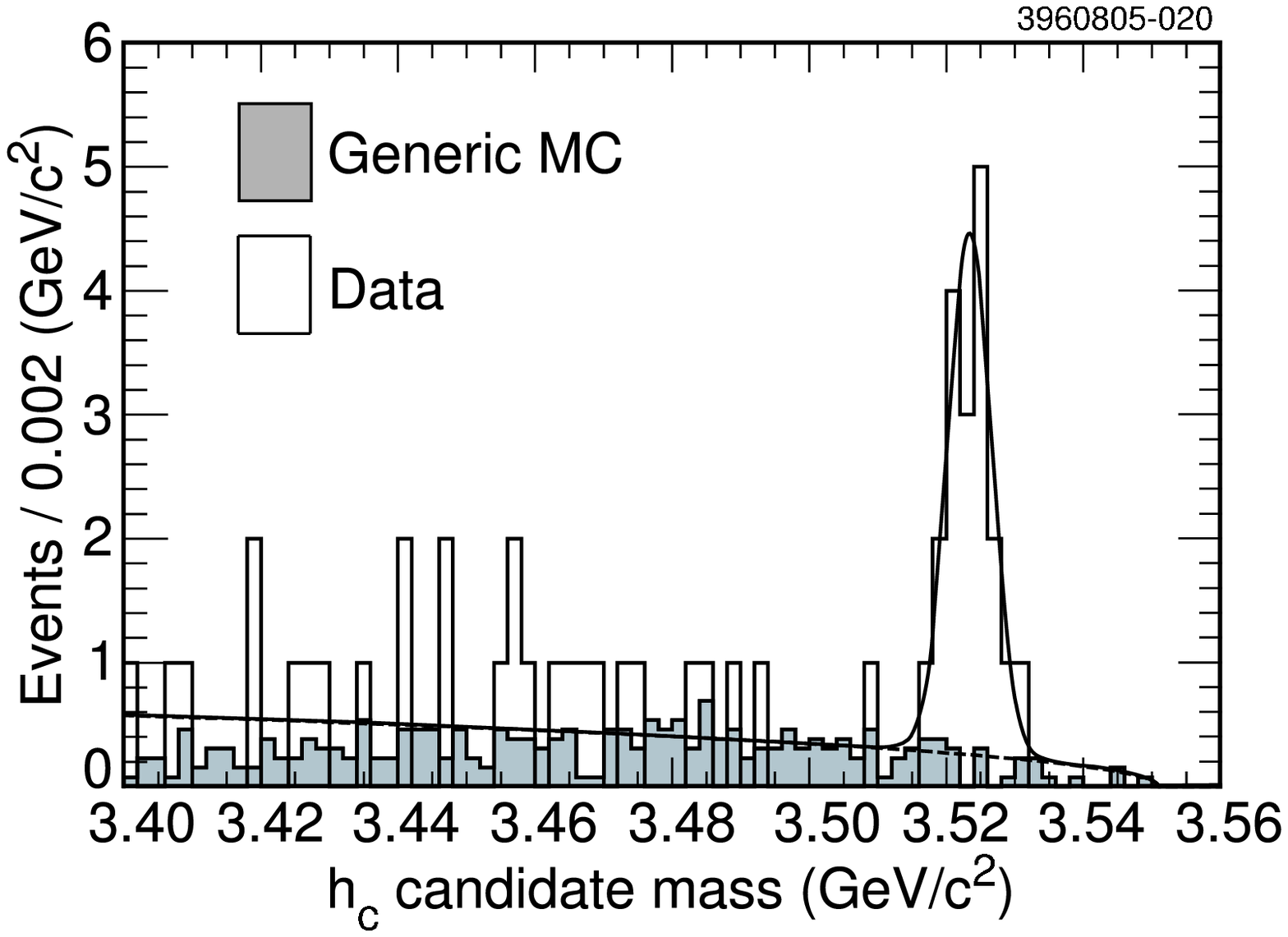,height=3.4in}
\caption{Fitted $\pi^{0}$ recoil mass of $\hc$ candidate for $M(\ec) = 2982 \pm
50$ MeV/$c^2$ in exclusive analysis.  Data events correspond to open histogram;
Monte Carlo background estimate is denoted by shaded histogram.
The signal shape is a double Gaussian, obtained from signal Monte Carlo.
The background shape is an ARGUS function.
\label{fig:fithcmass}}
\end{center}

\begin{center}
\vskip 0.1in
\epsfig{file=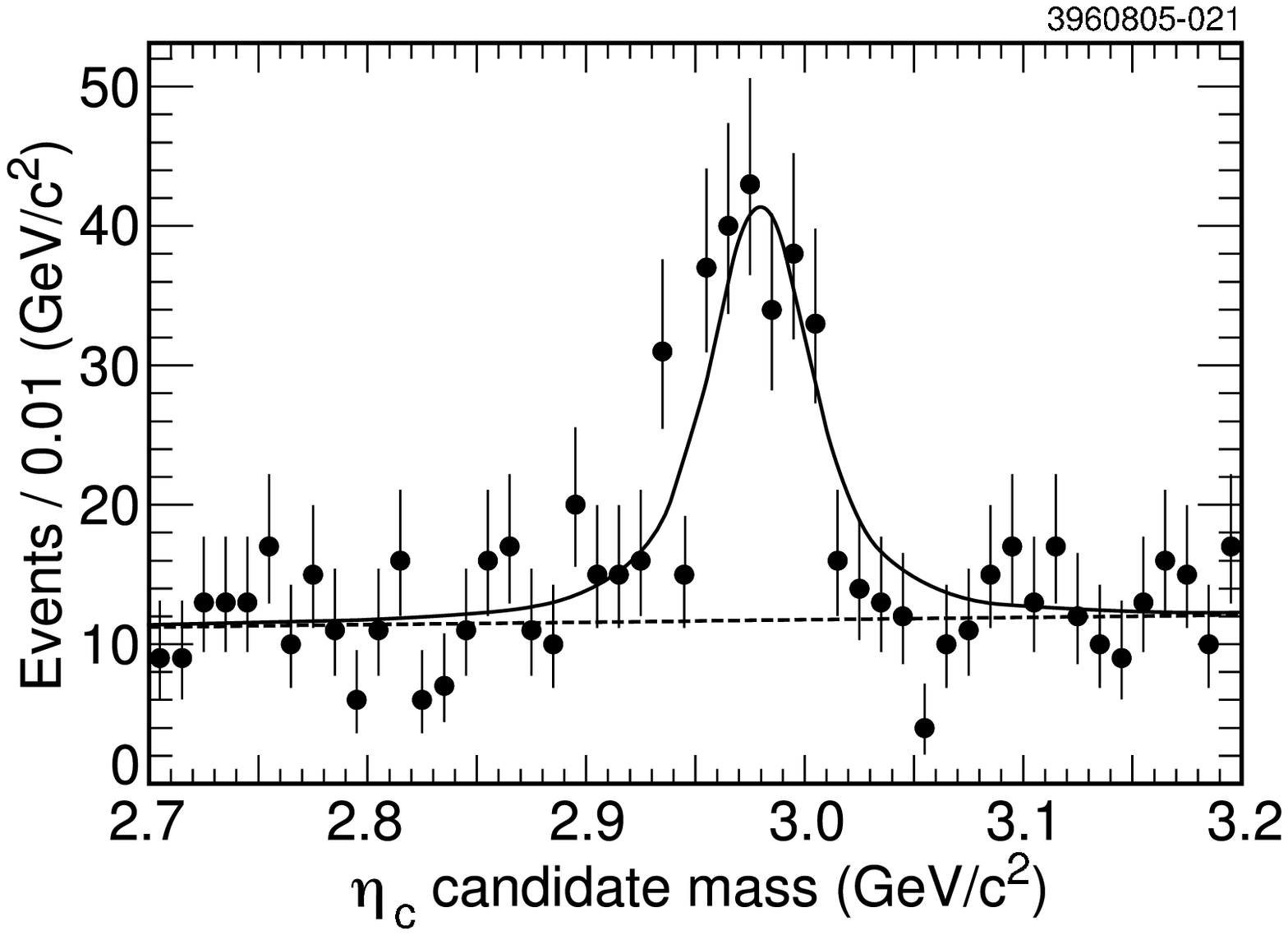,height=3.4in}
\caption{Fitted photon recoil mass in data ($\direct$, exclusive analysis).
The signal shape is
a double Gaussian convolved with a Breit-Wigner function. The mass resolution
function is obtained from signal Monte Carlo. The background shape is a
first-order polynomial function. The $\eta_{c}$ mass is fixed at the value
\cite{Eidelman:2004wy} 2979.7 MeV/$c^2$.
\label{fig:radetacmass}}
\end{center}
\end{figure}

There is a highly populated band at the $J/\psi$ mass in Fig.\
\ref{fig:hcetacmass}.  Monte Carlo studies indicate that most of these
events are from $\pi^{0}\pi^{0}J/\psi$ and $\gamma\chi_{cJ}(J=0,1,2)$.  When
one soft photon from a $\pz$ of $\pi^{0}\pi^{0}J/\psi$ is missing, neither the
beam energy constraint nor $\pi^{0}$ suppression can remove this background,
but $\eta_{c}$ mass selection is powerful in rejecting such events.  Once this
selection is imposed, corresponding to the range $M(\ec) = 2982 \pm 50$
MeV/$c^2$ in Fig.\ \ref{fig:etacmass}, a clearer $h_{c}$ signal appears in the
$\pi^{0}$ recoil mass spectrum around 3525 MeV/$c^2$ (Fig.\
\ref{fig:fithcmass}).
The distribution was fitted using an unbinned maximum likelihood method and
ARGUS background function to obtain the yield and the mass of the observed
$h_{c}$ signal.  The double Gaussian signal shape is obtained from signal Monte
Carlo in which the dominant narrower Gaussian width is 3.2 MeV/$c^2$.  The
unbinned maximum likelihood fit yields $17.5 \pm 4.5$ $\hc$ candidates with
mass at $3523.6 \pm 0.9$ MeV/$c^2$.  The
significance of the signal calculated from the difference in the likelihood
with and without the signal contribution is $6.1 \sigma$.

A clear $\eta_{c}$ signal also is observed in mass recoiling against the photon
in the study of the radiative
decay $\direct$. This confirms the appropriateness and
effectiveness of the event selection criteria. The recoil mass resolution is
identical for all modes, and independent of track momentum resolution.
The signal shape function, a Breit-Wigner function convolved with a double
Gaussian, is obtained from signal Monte Carlo. The width of the Breit-Wigner
function represents the $\eta_{c}$ intrinsic width. The detector resolution,
represented by a double Gaussian, was obtained by fitting the distribution of
the difference between the generated and reconstructed $\eta_{c}$ candidate
masses.

A total of $220 \pm 22$ events in all seven modes was observed
(Fig.~\ref{fig:radetacmass}).  The ratio of the branching ratios $\br$ for the
cascade ($\cascade$) and direct radiative ($\direct$) decays in each mode is
shown in Table \ref{tb:brratio}.  To calculate the resulting event-weighted
average ratio $\bhc/\bdir \equiv {\cal B}(\cascade)/{\cal B}(\direct)$,
one may write the observed number $N(X,\hc)$ of $\ec$ decays via $\cascade$ and
the observed number $N(X,{\rm dir})$ via $\direct$ to an $\ec$ channel $X$ with
${\cal B}(\ec \to X) \equiv {\cal B}(X)$ respectively as
\beq
N(X,\hc) = \bhc {\cal B}(X) N(\pp) \epsilon(X,\hc)~~,~~~
N(X,{\rm dir}) = \bdir {\cal B}(X) N(\pp) \epsilon(X,{\rm dir}),
\eeq
where $\epsilon(X,{\rm dir})$ and $\epsilon(X,\hc)$ are efficiencies for mode
$X$ for direct and cascade decays (Table \ref{tb:brratio}).  One then finds
\beq
\frac{\bhc}{\bdir} = \frac{\sum_X N(X,\hc)}{\sum_X N(X,{\rm dir})}/
\frac{\sum_X \epsilon(X,\hc){\cal B}(X)}{\sum_X \epsilon(X,{\rm dir})
{\cal B}(X)} = 0.178 \pm 0.049~(\stat),
\eeq
where $\sum_X N(X,\hc) = 17.5 \pm 4.5$ and $\sum_X N(X,{\rm dir})=220 \pm 22$.
\bigskip

\begin{table}
\caption{Efficiencies and yields of direct radiative decay ($\direct$) and
cascade decay ($\cascade$) in exclusive analysis, and ratio of branching
ratios, for each mode.} \label{tb:brratio}
\begin{center}
\begin{tabular}{|l|c|c|c|c|c| }
\hline
 & \multicolumn{2}{|c|}{ direct radiative decay} & \multicolumn{2}{|c|}
{ cascade decay} & ${\cal B}$(cascade)/ \\ \cline{2-5}
 \raisebox{1.5ex}[0pt]{Mode} &Eff(\%)&Yield&Eff(\%) & Yield &
 ${\cal B}$(direct) \\ \hline
$\kskp$ &12.7&$35.5\pm7.6$&5.6&1.9$\pm$1.4&0.116$\pm$0.090\\ \hline
$\klkp$ &32.6&74.0$\pm$12.0&15.3&3.1$\pm$2.1&0.081$\pm$0.057 \\ \hline
$\kkpp$ & 24.9 & 10.3$\pm$6.9 & 10.8 & 2.8$\pm$1.7 & 0.633$\pm$0.673\\ \hline
$\fourpi$ & 35.6 & 46.0 $\pm$12.0 & 15.1 &7.3$\pm$2.8 & 0.290$\pm$0.132\\
\hline
$\kkpizero$ & 24.2 & 21.6$\pm$6.4 & 10.9 & 0.9$\pm$1.0 & 0.098$\pm$0.114\\
\hline
$\ppetaone$ & 30.6  & 23.7$\pm$6.9&14.8&0.0+1.0%
\footnote{We estimate the error of the yield to be 1 according to
the Poisson distribution.} &0.000+0.083\\
\hline
$\ppetatwo$ &16.4&12.7$\pm$4.8&7.3&1.0$\pm$1.0&0.205$\pm$0.225\\ \hline
Total & - &220$\pm$22&-&17.5$\pm$4.5&0.178$\pm$0.049\\ \hline
\end{tabular}
\end{center}
\end{table}

\clearpage
\centerline{\bf VIII.  SYSTEMATIC ERRORS}
\bigskip

The systematic errors on $M(\hc)$ and $\brp \brh$ are summarized in Table
\ref{tab:sys}.  The following subsections describe how these errors
were obtained in the individual analyses.  When different approaches yield
different results, the most conservative value is entered.

\begin{table}[h]
\caption{Comparison of systematic errors in $M(\hc)$ and $\brp \brh$ for
inclusive and exclusive analyses.  N/A:  not applicable.
\label{tab:sys}}
\begin{center}
\begin{tabular}{|l|c c|c c|} \hline \hline
 & \multicolumn{2}{c|}{$M(h_c)$, MeV/$c^2$}
 & \multicolumn{2}{c|}{${\cal B}_1 \times {\cal B}_2 \times 10^4$} \\
Systematics in & Inclusive & Exclusive & Inclusive & Exclusive \\ \hline
Number of $\pp$ & N/A & N/A & 0.1 & N/A \\
${\cal B}(\pp \to \gamma \eta_c)$ & N/A & N/A & N/A & 0.8 \\ \hline
Background shape & 0.3 & 0.2 & 0.2 & 0.3 \\
$\pi^0$ energy scale & 0.2 & 0.2 & $\sim 0$ & 0.1 \\
Signal shape & 0.1 & 0.1 & 0.3 & 0.2 \\
$h_c$ width & 0.1 & 0.1 & 0.3 & 0.2 \\
$\pi^0$ efficiency & $\sim 0$ & $\sim 0$ & 0.2 & 0.3 \\
E1 Photon efficiency & $\sim 0$ & $\sim 0$ & 0.2 & 0.2 \\ 
Binning, fitting range & 0.1 & 0.1 & 0.3 & 0.2 \\
Modeling of $h_c$ decays & 0.1 & 0.3 & 0.3 & $\sim 0$ \\
$\eta_c$ mass & 0.1 & 0.2 & 0.1 & 0.1 \\
$\eta_c$ width & $\sim 0$ & $\sim 0$ & 0.2 & 0.1 \\
$\eta_c$ branching ratios & N/A & $\sim 0$ & N/A & 0.1 \\ \hline
Sum in quadrature & $\pm0.4$ & $\pm0.5$ & $\pm0.7$ & $\pm1.0$ \\ \hline \hline
\end{tabular}
\end{center}
\end{table}

\bigskip
\leftline{\bf A.  Inclusive analyses}
\bigskip

\leftline{\it 1.  Choice of background.}

Final results in the $\ege1$ analysis were obtained using the $\pi^0$ recoil
background generated from the data.  To estimate the systematic error due to
choice of background, data were also fitted with a generic Monte Carlo
background shape, yielding systematic uncertainties $\Delta M(\hc) \sim 0.2$
MeV/$c^2$, $\Delta \brp \brh \sim 0.2\times10^{-4}$.  A similar value of
$\Delta \brp \brh$ was obtained in the $M(\ec)$ analysis by replacing generic
Monte Carlo background by a second order polynomial plus an ARGUS function.
However, a slightly larger value of $\Delta M(\hc) \sim 0.3 \times
10^{-4}$ was seen both in data and in Monte Carlo.  It is this value we quote
in Table \ref{tab:sys}.

\bigskip
\leftline{\it 2.  Photon energy calibration for $\pz$ energy scale.}



The standard CLEO CsI calorimeter calibration was used.  To determine if the
uncertainty in this calibration can lead to systematic error in $E(\pz)$,
the total deposited calorimeter energy was varied by amounts estimated by
studies of radiative transitions in $\psi(2S)$ \cite{Athar:2004dn} and
$\pi^0 \to \gamma \gamma$ found in data.  The analysis procedure, including
fitting, was then repeated with Monte Carlo data to check for dependence on
absolute calibration of CC energy.  The small effects found may be ascribed in
part to the compensating effect of the demand that the two photons in the
low-energy $\pz$ have the correct effective mass.  We assign an error of $\pm
0.2$ MeV/$c^2$ in $M(\hc)$ to the $\pz$ energy scale on the basis of the
arguments advanced in the subsection on the exclusive analysis.

\bigskip

\bigskip
\leftline{\it 3. Signal shape.}

The systematic uncertainty due to uncertainty in the $\pi^0$ line shape was
found
by varying the Gaussian part of the signal shape by 10\% to account for a
possible mis-modeling (via Monte Carlo) of photon energy resolution
to be $\sim0.1$ MeV/$c^2$ in $M(h_c)$,
and $\sim 0.3\times10^{-4}$ in $\brp \brh$.

\bigskip
\leftline{\it 4.  Choice of $h_c$ resonance width.}

The systematic uncertainty due to variation of $\Gamma(h_{c})$ (0.5, 0.9, 1.5
MeV) was found to be $\sim$0.1 MeV/$c^2$ in $M(h_c)$, and $\sim 0.3
\times10^{-4}$ in $\brp \brh$.  Variation of the Gaussian widths by $\pm10\%$
led to negligible changes in mass and combined branching ratio.

\bigskip
\leftline{\it 5.  Binning and fitting range.}

In the $\ege1$ analysis the systematic uncertainty due to fit using 1 MeV/$c^2$
bins, instead of the usual 2 MeV/$c^2$ bins, and changing the fitting range
from 3496--3552 MeV/$c^2$ to 3500--3540 MeV/$c^2$ (see Table \ref{tab:summ})
was found to be $\le 0.1$ MeV/$c^2$ in $M(h_c)$, and $\le 0.2 \times 10^{-4}$
in $\brp \brh$.  The $M(\ec)$ analysis chose 1 MeV/$c^2$ bins to utilize the
good $M(\hc)$ resolution anticipated from Monte Carlo simulations.  Results
were compared with those from 2 MeV/$c^2$ bins and agreed with those just
quoted.  For the fitting range 3505-3551.2 MeV/$c^2$ in this analysis, however,
$\brp \brh$ in data rose by $0.3 \times 10^{-4}$.  This change was included
as a systematic error associated with fitting.

\bigskip
\leftline{\it 6.  Modeling of $h_c$ decays.}

The signal Monte Carlo used in the $\ege1$ analysis took 100\% of $h_c$
decaying to to $\gamma\eta_c$.  An alternative signal Monte Carlo, in which
37.7\% of $h_c$ were taken to decay to $\gamma \eta_c$ and the rest to
three gluons was generated and used to redetermine efficiency.  The resulting
$\brp \brh$ changed by $\sim 0.1 \times 10^{-4}$.  However, in the $M(\ec)$
analysis, larger differences were observed in Monte Carlo simulations when
comparing $\br(\hc \to \gamma \ec) = 37.7\%, \br(\hc \to ggg) = 56.8\%,
\br(\hc \to \gamma gg) = 5.5\%$ (nominal) and $\br(h_c \to \gamma \ec) =
100\%$.  The nominal choice gave about 10\% higher efficiency since events of
the form $\pp \to \pz \hc$ with $\hc \to ggg$ or $\hc \to \gamma g g$ sometimes
pass signal selection criteria.  The systematic error of $0.3 \times 10^{-4}$
quoted in Table \ref{tab:sys} reflects this larger value.

\bigskip
\leftline{\it 7.  Selected $M(\eta_c)$ range.}

In the $M(\ec)$ inclusive analysis, the 13 small Monte Carlo samples show that
neither $M(\hc)$ nor $\brp \brh$ is very sensitive to the selected $M(\ec)$
range in the intervals 2940--3020, 2945--3015, 2950--3010, 2955--3005, and
2960--3000 MeV/$c^2$, leading to errors of $\pm 0.1$ MeV/$c^2$ in $M(\hc)$ and
$\pm 0.1 \times 10^{-4}$ in $\brp \brh$.

\bigskip

\leftline{\it 8.  Removal of ``pull mass'' requirement on signal $\pi^0$.}

Instead of requiring that the signal $\pi^0$ possess the best ``pull mass''
within $2.5 \sigma$, {\it all} two-photon combinations with $M(\pz)^2$ within
$2.5 \sigma$ of the correct value were considered in the $M(\ec)$ analysis.
The maximum signal
significance as measured by likelihood difference in Monte Carlo was reduced
from $17.3 \sigma$ (Table \ref{tab:MCcomp}) to $16.1 \sigma$ for the nominal
$M(\ec)$ range 2945--3015 MeV/$c^2$.  Although $M(\hc)$ obtained in the data
shifted by $+0.1$ MeV/$c^2$ from the nominal value, while the branching ratio
shifted by $+0.9 \times 10^{-4}$ from the nominal value, these shifts are
within the statistical errors.  No such shifts were detected in Monte Carlo
simulations.  Consequently, systematic errors were assigned to the effect of
removing the pull mass requirement on the signal $\pz$ of less than
$0.1$ MeV/$c^2$ in $M(\hc)$ and $0.1 \times 10^{-4}$ in $\brp \brh$.
\bigskip

\leftline{\it 9.  Number of neutral pions in signal region.}

Both inclusive analyses require that there be only one $\pz$ candidate yielding
a recoil
$\hc$ mass within 30 MeV/$c^2$ of 3526 MeV/$c^2$.  The effect of relaxing
this condition was noted.  In all cases (independently of other selection
choices), it led to Monte Carlo significances which decreased by 0.2--$0.3
\sigma$, a decrease of $M(\hc)$ by about 0.1 MeV/$c^2$ and $\brp \brh$
by $0.3 \times 10^{-4}$ in data, but negligible changes in $M(\hc)$ and
$\brp \brh$ in Monte Carlo.  Systematic errors in $M(\hc)$ and $\brp \brh$
from this source were estimated to be less than $\pm 0.1$
MeV/$c^2$ and $\pm 0.1 \times 10^{4}$, respectively.
\bigskip

\leftline{\it 10.  Mass ranges for $\pp \to X J/\psi$ cascade suppression.}

In the $M(\ec)$ analysis, nominal mass ranges to suppress
$\pi^+ \pi^- J/\psi$, $\pi^0 \pi^0 J/\psi$, and $\gamma \gamma J/\psi$ cascades
involve recoil masses differing from $M(J/\psi)$ respectively by 8.4 MeV/$c^2$
($\pi^+ \pi^-$), 32 MeV/$c^2$ ($\pi^0 \pi^0$), and 40 MeV/$c^2$ ($\gamma
\gamma$).  These values were varied over the respective ranges 6.4--10.4,
22--42, and 30--50 MeV/$c^2$.  The maximum variations from each mode were then
added in quadrature.  Possible changes of $\pm 0.2$ MeV/$c^2$ in $M(\hc)$ and
$\pm 0.2\times 10^{-4}$ in $\brp \brh$ were seen in data, but negligible
changes occurred in Monte Carlo simulations.  These sources were thus estimated
to lead to systematic errors of $\Delta M(\hc) < 0.1$ MeV/$c^2$ and $\Delta
\brp \brh < 0.1 \times 10^{-4}$.
\bigskip

\leftline{\it 11.  Minimum energy requirements on photons.}

In suppressing $\pz \pz J/\psi$ cascades, a minimum energy of 50 MeV was taken
for photon daughters in the $M(\ec)$ analysis.  The result of reducing this
energy to 40 MeV was a
stronger suppression of both background and signal, leading to an upward shift
of the mass by 0.2 MeV/$c^2$ in data and no change in $\brp \brh$ in data.
Changes in mass and $\brp \brh$ were negligible in Monte Carlo.
\bigskip

\leftline{\it 12.  Correction for updated $M(\pp)$.}

The $M(\ec)$ analysis was based on the assumption of $M(\pp)=3685.96 \pm 0.09$
MeV/$c^2$, the world average \cite{Hagiwara:2002fs} before the measurement of
Ref.\ \cite{Aulchenko:2003qq}.  With the present value of $M(\pp) = 3686.111
\pm0.025 \pm0.009$ MeV/$c^2$, a correction of $+0.15$ MeV/$c^2$ thus was
applied to the final quoted mass in that analysis.
\bigskip

\leftline{\bf B.  Exclusive analysis}
\bigskip

Because the exclusive cascade rates were measured as ratios to the radiative
decays, systematic uncertainties related to the $\eta_c$~final state cancel.
The systematic studies dealt with estimating the statistical significance of
the $h_{c}$ signal, the $h_{c}$ mass, and the production branching ratio.

In order to study the background contribution from the non-$\pp$ part of the
data (continuum data), 22 pb$^{-1}$ of continuum data (beam energy $\simeq 
1835~{\rm MeV} = M(\pp)/2 - 7.5$ MeV) were analyzed in the same manner. The
contribution of continuum data was found to be negligible.

The generic Monte Carlo sample was used to see if any of the known $\pp$ decays
could produce a fake peak which would mimic the signal.  No significant peak
was seen in the signal region (8 bins in the $\pi^{0}$ recoil mass histogram,
from 3516 to 3532 MeV/$c^2$) with 39 million generic Monte Carlo events
(13 times the data sample).  This implies the
signal seen in data is not due to a reflection of any known charmonium decays.

The significance can be estimated from the background level in the signal
region using the generic Monte Carlo or data sideband.
Using events from the likelihood values of the fit with and without the
signal contribution, we obtain $s = 6.1 \sigma$; similar calculations with
different $\eta_c$ mass ranges yield $s=5.5-6.6\sigma$.  Using events from
the generic Monte Carlo sample, appropriately scaled so as to match event
populations outside the signal region, we obtain an estimate of a mean
background inside the signal window of $2.5 \pm 0.5$ events.  Allowing for
Poisson fluctuations of this number results in a probability that background
completely accounts for the observed signal of 19 events of $1 \times 10^{-9}$
($s = 6.0 \sigma$).  The binomial probability that the 47 data
events in Fig.\ \ref{fig:fithcmass} and the 8 data events
in the $\eta_c$ sideband, $2600 \le M(h_c) \le 2860$ MeV/$c^2$, of Fig.\
\ref{fig:etacmass} fluctuate to be greater than the 19 events
in the signal region, $3516 < M(h_c) < 3532$ MeV/$c^2$, of Fig.\
\ref{fig:fithcmass} is $2.2 \times 10^{-7}$, which corresponds to a
significance of $\sim 5.2 \sigma$.
Estimates of signal significance are summarized in Table \ref{tb:chksys}.

\begin{table} 
\caption{Checks of significance, $h_{c}$ mass and production branching ratio
($\br(\cascade)$ stability by varying key selection criteria (exclusive
analysis).} \label{tb:chksys}
\begin{center}
\begin{tabular}{|l|c|c|c| } \hline \hline
Selection & Mass & ${\cal B}$(cascade)/ & Significance \\
        & (MeV/$c^2$) & ${\cal B}$(direct) & $(\sigma)$ \\ \hline
\textbf{Default cuts} & \textbf{3523.6$\pm$0.9} & \textbf{0.178$\pm$0.049} &
 \textbf{6.1} \\ \hline
Fit $\chi^{2} < 3$ & +0.1&0.192$\pm$0.056 & 6.2\\ \hline
Fit $\chi^{2} < 5$ & +0.5& 0.178$\pm$0.051& 6.1\\ \hline
Fit $\chi^{2} <$ 15 & 0.0&0.169$\pm$0.049 & 5.8\\ \hline
Within 30 $\MeV$ of $\eta_{c}$ mass&+0.7&0.165$\pm$0.50&5.5\\ \hline
Within 40 $\MeV$ of $\eta_{c}$ mass&+0.2&0.172$\pm$0.049&5.9\\ \hline
Within 60 $\MeV$ of $\eta_{c}$ mass&0.0&0.172$\pm$0.049&5.9\\ \hline
Within 80 $\MeV$ of $\eta_{c}$ mass&-0.1&0.188$\pm$0.052&6.6\\ \hline
Transition photon $\pi^{0}$ veto (2$\sigma$)&0.0&0.168$\pm$0.051&5.6\\ \hline
Transition photon $\pi^{0}$ veto (4$\sigma$)&+0.2&0.152$\pm$0.046&5.9\\ \hline
Kinematic fitted $h_{c}$ &+0.4&0.166$\pm$0.049&5.6\\ \hline
CLEOIII only &+0.5&0.158$\pm$0.069&3.9\\ \hline
CLEOc only &-0.3&0.216$\pm$0.083&4.7\\ \hline \hline
\end{tabular} \\
\end{center}
\end{table}

The mass of $\hc$ is estimated from a $\pi^{0}$ recoil mass calculation.  The
systematic uncertainty associated with this estimate depends on the uncertainty
of the $\pi^{0}$ energy scale, which is itself dependent on the energies of the
photon daughters and their shower locations in the detector.  Lower-energy
photons and endcap photons have larger associated uncertainties.
The fraction of endcap photons is small ($<$10\%), so the
shower-location effect on energy resolution was ignored.  The signal $\pi^0$
energy is around 160 MeV, and the corresponding $\pz$ daughter photon energies
vary from 30 to 130 MeV, with respective uncertainties varying from 1.5\% to
0.2\%.  By changing the
photon energy uniformly by $\pm$1\%, the $\pi^{0}$ energy in the signal Monte
Carlo was found to shift only less than $\pm 0.2$ MeV because of the $\pi^{0}$
mass constraint in the analysis algorithm which fits neutral pions.
Consequently, a 0.2 MeV systematic uncertainty in $M(\hc)$ was ascribed to the
$\pi^{0}$ energy scale.

 
The $\ec$ intrinsic width $\gec$ has not been
accurately measured.  In the exclusive signal Monte Carlo, it is set at 27 MeV.
Because the efficiency for detecting $h_{c}$ is estimated from signal Monte
Carlo and a range of $M(\ec)$ is selected, an overestimate of $\gec$ will
result in an underestimated efficiency. On the other hand, it will lead to a
wider signal shape for the $\ec$ signal in $\direct$ and hence to an increased
$\ec$ yield.  Thus the systematic error on the measured ratio of rates for
$\cascade$ and $\direct$ is likely to be small because the two effects tend to
cancel each other.  A 2.3\% systematic error was assigned to the ratio from the uncertainty in the $\eta_{c}$ intrinsic
width.

The uncertainties in the $\eta_{c}$ decay branching ratios are large; no
channel is known to better than 25\%.  Changing the branching ratio of each
mode 40\%, once per mode, the measured ratio was found to shift less than 1\%.
Consequently, a 1\% systematic error on the ratio of rates was ascribed to
$\eta_{c}$ decay branching ratios uncertainties.

In the analysis of the photon recoil mass from the direct radiative decay, the
$\eta_{c}$ mass was fixed at $2979.7$ MeV/$c^2$.  When this mass was floated
in fitting, the value determined from the fit was $M(\ec) = 2970.3 \pm 4.1$
MeV/$c^2$. This result is lower than, but still consistent with, the CLEO
inclusive photon transition study, in which the measured $\eta_{c}$ mass
is $2976.1 \pm 2.3 \pm 3.3$ \cite{Skwarnicki:2003wn}.  Varying the fixed value
of the $\ec$ mass in the fit of the recoil mass distribution between 2970 and
2984 MeV/$c^2$ resulted in a variation of 3.6\% in the yield.  Half of this
value, 1.8\%, was assigned to the systematic uncertainty of the combined
branching ratio due to uncertainty in $M(\ec)$.

In the decay $\cascade$, the $\eta_c$ mass selection is based on the value
obtained by reconstructing the $\eta_c$.  When the $\eta_c$ mass selection
window is shifted by $\pm$ 10 MeV/$c^2$, the measured value of $M(\hc)$ shifts
by less than 0.2 MeV/$c^2$.  We assign 0.2 MeV/$c^2$ as the $h_{c}$ mass
systematic uncertainty due to uncertainty in $M(\ec)$.

Neutral pion reconstruction efficiency has been studied in
measurements of $D$ hadronic branching fractions. The discrepancy between Monte
Carlo and data is less than 5\% \cite{He:2005bs}.  We ascribe a 5\% systematic
uncertainty in the ratio of rates to $\pi^{0}$ efficiency uncertainty.
This corresponds to an uncertainty in the product branching ratio of $0.27
\times 10^{-4}$ for the exclusive analysis and $0.18 \times 10^{-4}$ for the
inclusive analysis (which finds a slightly smaller product branching ratio).

In the signal Monte Carlo for the exclusive analysis, $\ghc$ was set to zero, so
the signal shape obtained from Monte Carlo essentially represented detector
resolution.  Varying the assumed value of $\ghc$ up to 1.5 MeV changed
the measured $h_{c}$ mass by less than 0.1 MeV/$c^2$ and the branching ratio
by 3.9\%.  We also studied the effects of the signal shape by changing detector
resolution by $\pm$ 20\%. The background in the exclusive study is quite small,
so the $\pi^{0}$ recoil mass fit range was chosen
starting from 3400 MeV$/c^2$.  The wider background range helped to fit the
background shape better.  Varying the starting point of the fit from 3.40 to
3480 MeV$/c^2$ did not change the mass and branching ratio measurement much.
First- and second-order polynomial background shapes were used to fit the
background and to study the systematics.  The mass change was 0.2 MeV/$c^2$ and
the rate change was 4.7\%.

The $\chi^{2}$ limit in kinematically constrained fits, the selection of the
range for $M(\eta_{c})$, and the veto of E1 transition photon candidates
forming a $\pi^{0}$ were found to be the most useful selection criteria in the
exclusive study.  Variation of these selection criteria within reasonable
ranges did not change
the corresponding $h_{c}$ mass and product branching ratios appreciably.
The resolution in $M(\hc)$ obtained using $\pi^0$ momentum after kinematic fits
was slightly better than that from measured $E(\pi^{0})$ by 2--5\%, depending
on modes.  Because different mass resolutions lead to difficulty in obtaining
results and the possible gain in the mass measurement is small, momentum
fitting was not used to obtain $M(\hc)$.  Using the kinematically fitted
$h_{c}$ mass yielded values of $M(\hc)$ and production branching ratio
consistent with nominal results.
\bigskip

\centerline{\bf IX.  SUMMARY AND DISCUSSION}
\bigskip

\leftline{\bf A.  Inclusive analyses: Summary}
\bigskip

Two inclusive analyses of CLEO data in search of
$\cascade$ yield an enhancement in the mass spectrum for recoils against
$\pi^0$ attributed to the $h_c(1^1P_1)$ resonance of charmonium.
When background is reduced by selecting a range of photon energies
$\ege1 = 503 \pm 35$ MeV, the parameters of the resonance are found to be
\beq
M(h_{c}) = [3524.4 \pm 0.7~(\stat) \pm 0.4~(\sys)]{\rm~MeV}/c^2,
\eeq
\beq
\brp \brh \equiv \br(\pp \to \pz \hc) \times \br(\hc \to \gamma \ec)
 = [3.4 \pm 1.0~(\stat) \pm 0.7~(\sys)] \times 10^{-4}.
\eeq
The significance of the resonance signal in this analysis, as determined by the
likelihood method, is $3.6\sigma$.  When background is reduced by selecting a
range of $M(\ec) \pm 35$ MeV/$c^2$, to compensate for Doppler broadening of the
photon in the transition $h_c \to \gamma \ec$ arising from the $\hc$ recoil,
one finds
\beq
M(\hc) = [3525.4 \pm 0.6~(\stat) \pm 0.4~(\sys)]{\rm~MeV}/c^2,
\eeq
\beq
\brp \brh = [3.5 \pm 0.9~(\stat) \pm 0.7~(\sys)] \times 10^{-4}.
\eeq
The significance of the resonance signal is $4.0 \sigma$.
\bigskip

\leftline{\bf B.  Exclusive analysis: Summary}
\bigskip

The $h_{c}$ produced in the reaction $\cascade$ was studied by reconstructing
$\eta_c$ in seven modes (Table \ref{tab:ecmodes}), leading to $17.5 \pm 4.5
(\stat)$ signal events. The significance as calculated from the difference in
the likelihood with and without the signal contribution is $6.1 \sigma$, and at
least $5.2 \sigma$ as calculated by a variety of methods.  The ratio of
$\br(\cascade)$ to $\br(\pp \to \gamma \ec)$ was found to be
\beq \label{eqn:ratio}
\frac{\br(\pp \to \pz \hc) \br(\hc \to \gamma \ec)}{\br(\pp \to \gamma \ec)}
 = 0.178 \pm 0.049~(\stat) \pm 0.018~(\sys),
\eeq
with
\beq
M(\hc) = [3523.6 \pm 0.9~(\stat) \pm 0.5~(\sys)] {\rm~MeV}/c^2.
\eeq

In CLEO III $\pp$ data, the branching ratio $\br(\direct)$ was measured to
be $(3.2 \pm 0.4~(\stat) \pm 0.6~(\sys)) \times 10^{-3}$ \cite{Athar:2004dn},
which when combined with previous measurements whose average is $(2.8 \pm 0.6)
\times 10^{-3}$ \cite{Eidelman:2004wy}, gives $\br(\direct) = (2.96 \pm 0.46)
\times 10^{-3}$.  Combining this with Eq.\ (\ref{eqn:ratio}), one obtains a
production branching ratio of
\beq
\brp \brh = [5.3 \pm 1.5~(\stat) \pm 0.6~({\rm internal~sys})
\pm 0.8~({\rm ext})]\times10^{-4},
\eeq
where the last error reflects the measurement error of $\br(\direct)$.  The
last two errors combine to give a total systematic error of $\Delta \brp \brh
= 1.0 \pm 10^{-4}$.
\bigskip

\leftline{\bf C.  Combination of results}
\bigskip

The results of the two inclusive analyses, when averaged (taking the larger
systematic and statistical errors in each analysis), yield $M(h_c) = [3524.9
\pm 0.7~(\stat) \pm 0.4~(\sys)$ MeV/$c^2$ and $\brp \brh=[3.5 \pm 1.0~(\stat)
\pm 0.7~(\sys)] \times 10^{-4}$.
The average is taken because, as explained in the second-to-last paragraph
of Sec.\ III, each inclusive analysis has its advantages and shortcomings,
without a clear preference for one over the other.
These results, which provide slightly more precise measurements of $M(h_c)$ and
$\brp \brh$, may be combined with the exclusive results, based on
reconstructing the $\ec$ in seven exclusive decay modes with much lower
background.
We have confirmed the independence of the exclusive analysis from the inclusive analyses by removing the exclusive signal events from our $\ege1$ inclusive
sample.  The results are indistinguishable from those of the original sample.
We therefore combine them to obtain
$M(\hc) = [3524.4 \pm 0.6~(\rm stat) \pm 0.4~({\rm sys})]$ MeV/$c^2$ and
$\brp \brh = [4.0 \pm 0.8~({\rm stat}) \pm 0.7~{\rm (sys)}] \times 10^{-4}$,
as summarized in Table \ref{tab:comb}.

\begin{table}
\caption{$M(h_c)$ and $\brp \brh$ obtained by the inclusive and exclusive
analyses; combined results.
\label{tab:comb}}
\begin{center}
\begin{tabular}{c c c} \hline \hline
Analysis     & $M(h_c)$ (MeV/$c^2$) & $\brp \brh$ (units of $10^{-4}$) \\
\hline
Inclusive $\ege1$  & $3524.4\pm0.7\pm0.4$ & $3.4 \pm 1.0 \pm 0.7$ \\
Inclusive $M(\ec)$ & $3525.4\pm0.6\pm0.4$ & $3.5 \pm 0.9^{+0.7}_{-0.4}$ \\
Avg.\ Inclusive    & $3524.9 \pm 0.7 \pm 0.4$ & $3.5 \pm 1.0 \pm 0.7$ \\
Exclusive    & $3523.6 \pm 0.9 \pm 0.4$ & $5.3 \pm 1.5 \pm 1.0$ \\
Incl.\ + Excl.\ & $3524.4 \pm 0.6 \pm 0.4$ & $4.0 \pm 0.8 \pm 0.7$ \\
\hline \hline
\end{tabular}
\end{center}
\end{table}
\bigskip

\leftline{\bf D.  Discussion}
\bigskip

The mass of the observed $\hc$ candidate is close to the spin-weighted average
of the $\chi_{cJ}$ states, $(3525.4 \pm 0.1)$ MeV/$c^2$.  This leads to
$\Delta M_{\rm HF}(1P)
\equiv \mav - M(1^1P_1) = [1.0 \pm 0.6~(\stat) \pm 0.4~(\sys)]$ MeV/$c^2$,
indicating little contribution of a long-range vector confining force or
coupled-channel effects which could cause a displacement from this value.
It is barely consistent with the (nonrelativistic) bound $\Delta M_{\rm HF}(1P)
\le 0$ \cite{Stubbe:1991qw}.  The product of the
branching ratios for its production, $\br(\pp \to \pz \hc)$, and its decay,
$\br(\hc \to \gamma \ec)$, is within the range anticipated theoretically.
\bigskip

\centerline{\bf ACKNOWLEDGMENTS}
\bigskip

We gratefully acknowledge the effort of the CESR staff 
in providing us with excellent luminosity and running conditions.
This work was supported by the National Science Foundation
and the United States Department of Energy. J. Rosner wishes to thank M. Tigner
for extending the hospitality of the Laboratory for Elementary-Particle
Physics at Cornell during part of this work and the John Simon Guggenheim
Foundation for partial support.

\end{document}